\RequirePackage{lineno}
\documentclass[aip,amsmath,amssymb,nofootinbib]{revtex4-1}
\usepackage{graphicx}
\usepackage{dcolumn}
\usepackage{bm}
\usepackage{color}
\usepackage{multirow}
\usepackage{colortbl}
\usepackage{subfig}
\usepackage{booktabs}
\usepackage{enumerate}
\usepackage{natbib}
\usepackage{hyperref}
\usepackage{listings}
\usepackage{color} 
\definecolor{mygreen}{RGB}{28,172,0} 
\definecolor{mylilas}{RGB}{170,55,241}

\begin{document}

\title{A Hybrid Profile--Gradient Approach for the Estimation of Surface Fluxes}

\author{Sukanta Basu}%
 \email{s.basu@tudelft.nl}
 \affiliation{Faculty of Civil Engineering and Geosciences, Delft University of Technology, Delft, the Netherlands}

\date{\today}

\begin{abstract}
The Monin--Obukhov similarity theory-based wind speed and potential temperature profiles are inherently coupled to each other. We have developed hybrid approaches to disentangle them, and as a direct consequence, the estimation of Obukhov length (and associated turbulent fluxes) from either wind-speed or temperature measurements becomes an effortless task. Additionally, our approaches give rise to two easily measurable indices of atmospheric stability. We compare these approaches with the traditional gradient and profile methods that require both wind-speed and temperature profile data.  Using Monte-Carlo-type numerical experiments we demonstrate that, if the input profiles are free of any random errors, the performance of the proposed hybrid approaches is almost equivalent to the profile method and better than the gradient method. However, the proposed hybrid approaches are less competitive in comparison to their traditional counterparts in the presence of random errors.   
\end{abstract}

\maketitle

\section{Introduction}
\label{Intro}

More than fifty years ago, in a classic paper, Panofsky~\cite{Panofsky63} wrote\footnote{The text within the parentheses, $[~]$, are included by Basu, S. to enhance readability.}: 
\begin{quotation}
``In principle, it should be possible to determine the three parameters $z_0$ [aerodynamic roughness length], $H$ [sensible heat flux], and $u_*$ [friction velocity] from three good wind observations close to the ground. But Priestkey~\cite{Priestley59} has pointed out that a small error in one or more of the winds leads to a huge error in the stress, so that this technique is not practical. Priestley further suggests that temperature data be added to the wind data in order that accurate estimates of stress be made. The present note considers this possibility in some detail.'' 
\end{quotation}
After this influential publication, the boundary-layer community at large embraced the idea and decided to focus on the estimation of turbulent fluxes utilizing both wind-speed and temperature data. The so-called gradient and profile methods (Appendix~1) were developed and refined. A few variants, using optimization techniques, were also proposed in parallel \citep[e.g.,][]{Nieuwstadt78}. 

In contrast, only a handful of studies did not follow suit. Swinbank~\cite{Swinbank64}, Klug~\cite{Klug67}, and Lo~\cite{Lo79} explored the possibility of estimating turbulent fluxes using only wind-speed measurements. Even though they documented reasonably good results, their flux-estimation approaches never received any serious attention in the literature. After all these years, it is difficult to pin-point the exact reasons behind their unpopularity. It is plausible that the inherent complexities of the approaches by Klug~\cite{Klug67} and Lo~\cite{Lo79} utilizing numerical optimization techniques rendered them less desirable in practical applications. Klug's approach also needed the aerodynamic roughness length ($z_0$) as an input, but accurate prescription of $z_0$ was (and still remains) a challenging task. The algorithm of Lo~\cite{Lo79} did not require $z_0$ as input, but suffered from convergence issues and possible mathematical errors \citep{Zhang81}. In addition, Lo~\cite{Lo79} did not include any error estimates of the derived variables as pointed out by Nieuwstadt and de Bruin~\cite{Nieuwstadt81}. The flux-estimation approach of Swinbank~\cite{Swinbank64} was more elegant, but was founded on the strong assumption that the surface-layer wind profile follows an exponential shape (Appendix~2). This assumption  departed significantly from the well-accepted logarithmic form (with correction terms) for the wind profile, which likely contributed to its unpopularity. 

With the advent of high-resolution, high-accuracy instruments for the measurement of wind speed and temperature (e.g., sodars, lidars, distributed temperature sensors), it is worthwhile to revisit the assertions made by Panofsky~\cite{Panofsky63}. The argument that we need both wind-speed and temperature measurements for flux estimation may no longer be tenable. At the same time, one needs to have a more analytically tractable approach than that advocated by Lo~\cite{Lo79} or Klug~\cite{Klug67}. Recently, in a short communication, we proposed such an approach, called the hybrid-wind approach \citep{Basu18}. With a few mathematical manipulations, we demonstrated that it is actually very straightforward to estimate turbulent fluxes from only wind-speed measurements. Our hybrid approach is similar to Swinbank~\cite{Swinbank64}. In the present study, we first extend this approach to utilize temperature data as input. Next, we compare the proposed hybrid approaches against traditional gradient and profile methods for a wide range of stability conditions. Last and most importantly, through uncertainty propagation experiments, we quantify the errors in estimated fluxes from all the aforementioned approaches. 

The structure of the paper is as follows: Sect.~\ref{Hybrid} introduces the newly proposed hybrid flux-estimation approach, and as by-products of this approach, two atmospheric stability indices are derived. Their characteristics are discussed in Sect.~\ref{Indices}. Some caveats of the proposed hybrid approaches are touched upon in Sect.~\ref{Limitations} and illustrative examples comparing the proposed approach and traditional flux-estimation approaches are documented in Sect.~\ref{Comparison}. The uncertainty propagation experiments and the associated results are also elaborated in this section. The concluding remarks including future directions are summarized in Sect.~\ref{Conclusion}. Background information on the traditional flux-estimation approaches, Swinbank's exponential wind-profile equation, and several relevant stability correction formulations are provided in the Appendices.

\section{Methodology}
\label{Hybrid}

The surface-layer wind speed and potential temperature profile equations based on the Monin--Obukhov similarity theory \citep[MOST;][]{Monin54} are written as,
\begin{subequations}
\begin{equation}
U\left(z\right) = \frac{u_*}{\kappa} \left[\ln\left(\frac{z}{z_0} \right) - \psi_m\left(\frac{z}{L}\right) + \psi_m\left(\frac{z_0}{L}\right)\right],
\label{MOST_U}
\end{equation}
\begin{equation}
\Theta\left(z\right) - \Theta_S = \frac{\theta_*}{\kappa} \left[\ln\left(\frac{z}{z_{0 T}} \right) - \psi_h\left(\frac{z}{L}\right) + \psi_h\left(\frac{z_{0 T}}{L}\right)\right],
\label{MOST_T}
\end{equation}
\end{subequations}
where, $\psi_m$ and $\psi_h$ are stability correction terms; $u_*$, $\theta_*$, and $L$ denote friction velocity, surface temperature scale, and Obukhov length, respectively. The aerodynamic roughness length and roughness length for temperature are represented by $z_0$ and $z_{0 T}$, respectively. $\Theta_S$ is the surface temperature, while the von K\'{a}rm\'{a}n constant is denoted by $\kappa$. 

Based on Eq.~\ref{MOST_U}, the vertical wind-speed difference (aka increment) can be computed as follows, 
\begin{subequations}
\begin{equation}
\Delta U_{21} = U\left(z_2\right) -  U\left(z_1\right) = \frac{u_*}{\kappa} \left[\ln\left(\frac{z_2}{z_1} \right) - \psi_m\left(\frac{z_2}{L}\right) + \psi_m\left(\frac{z_1}{L}\right)\right],
\label{dU1}
\end{equation}
\begin{equation}
\Delta U_{31} = U\left(z_3\right) -  U\left(z_1\right) = \frac{u_*}{\kappa} \left[\ln\left(\frac{z_3}{z_1} \right) - \psi_m\left(\frac{z_3}{L}\right) + \psi_m\left(\frac{z_1}{L}\right)\right], 
\label{dU2}
\end{equation}
\end{subequations}
where, $z_1$, $z_2$, and $z_3$, are the heights at which wind speed is measured. 

Finally, a ratio of the wind-speed differences can be written as, 
\begin{equation}
R_W = \frac{\Delta U_{31}}{\Delta U_{21}} = \frac{\ln\left(\frac{z_3}{z_1} \right) - \psi_m\left(\frac{z_3}{L}\right) + \psi_m\left(\frac{z_1}{L}\right)}{\ln\left(\frac{z_2}{z_1} \right) - \psi_m\left(\frac{z_2}{L}\right) + \psi_m\left(\frac{z_1}{L}\right)}.
\label{RatioW}
\end{equation}
In an analogous manner, a ratio of the potential temperature differences can be written as, 
\begin{equation}
R_T = \frac{\Delta \Theta_{31}}{\Delta \Theta_{21}} = \frac{\ln\left(\frac{z_3}{z_1} \right) - \psi_h\left(\frac{z_3}{L}\right) + \psi_h\left(\frac{z_1}{L}\right)}{\ln\left(\frac{z_2}{z_1} \right) - \psi_h\left(\frac{z_2}{L}\right) + \psi_h\left(\frac{z_1}{L}\right)}.
\label{RatioT}
\end{equation}
We strongly emphasize that the estimation of $R_W$ only requires observed wind-speed data from three levels; similarly, $R_T$ is solely based on temperature measurements at three levels. Due to their explicit functional relationships with $L$, both these quantities can be considered as independent proxies of atmospheric stability. In other words, both the wind-speed and temperature profile data are not required for the estimations of $L$ and associated fluxes; only one type of variable suffices. Illustrative examples are provided in Sect.~\ref{Comparison}. 

We have named our flux-estimation methodology a `hybrid' profile--gradient approach because it borrows ideas from both the traditional profile and gradient methods. Via mathematical manipulations, it  disentangles the original MOST equations, which has not been feasible in the traditional methods. Hereafter, we make a further distinction and refer to the proposed approach as `hybrid-W' or `hybrid-T' depending on whether wind-speed or temperature data are being utilized as inputs. 

\section{Characteristics of $R_W$ and $R_T$}
\label{Indices}

The behaviour of $R_W$ and $R_T$ depend entirely on the stability correction terms ($\psi_m$ and $\psi_h$). For neutral condition (i.e., $z/L = 0$), $\psi_m = \psi_h = 0$, whence, both ratios simplify to, 
\begin{equation}
R_N = \frac{\ln\left(\frac{z_3}{z_1} \right)}{\ln\left(\frac{z_2}{z_1} \right)}.
\label{RatioN}
\end{equation} 
If $z_3 > z_2 > z_1$, it is trivial to show that $R_N > 1$. Next, we consider the behaviour of $R_W$ and $R_T$  for non-neutral conditions. 

\begin{figure*}[ht]
\centering
\includegraphics[width=0.49\textwidth]{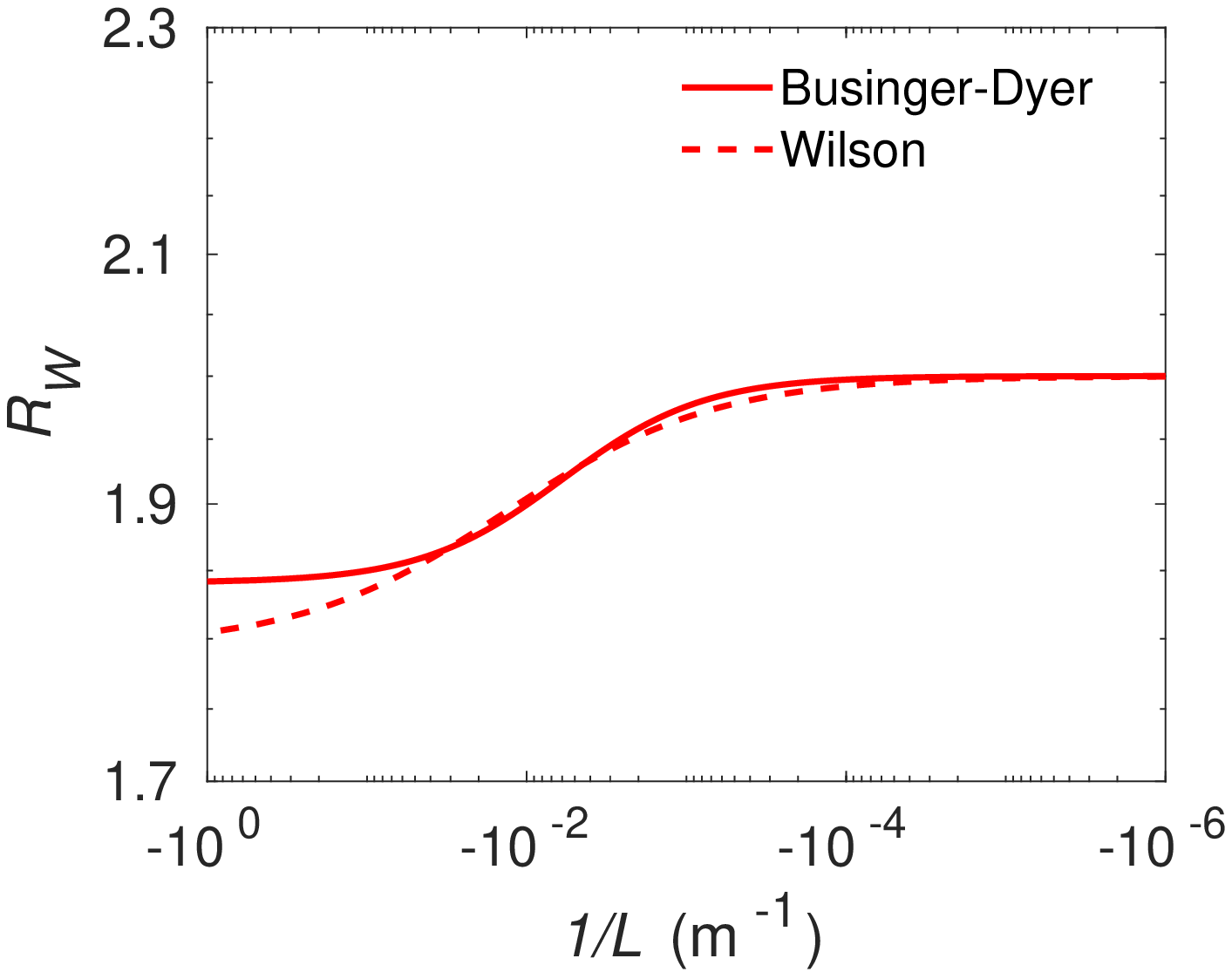}
\includegraphics[width=0.49\textwidth]{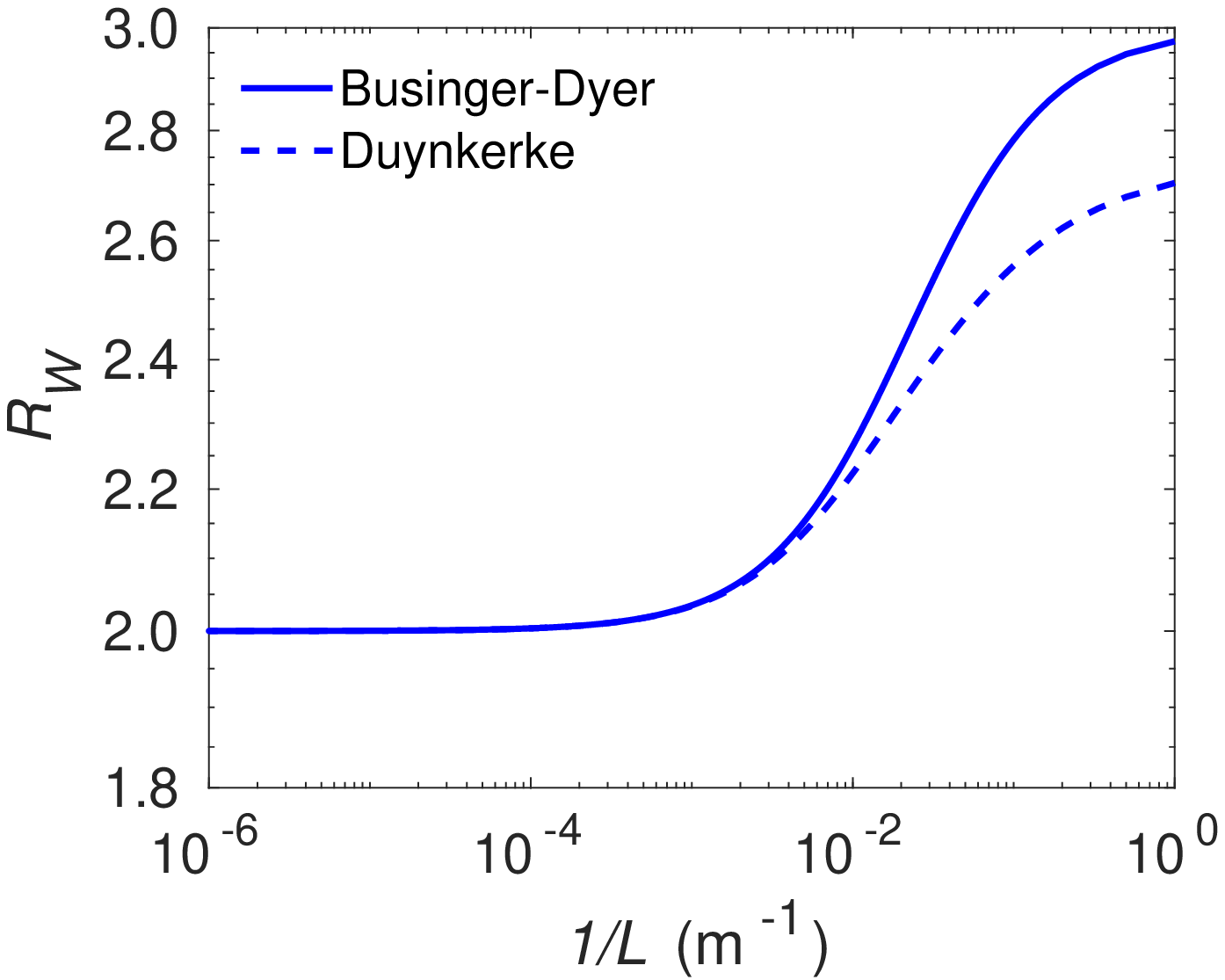}\\
\vspace{0.1in}
\includegraphics[width=0.49\textwidth]{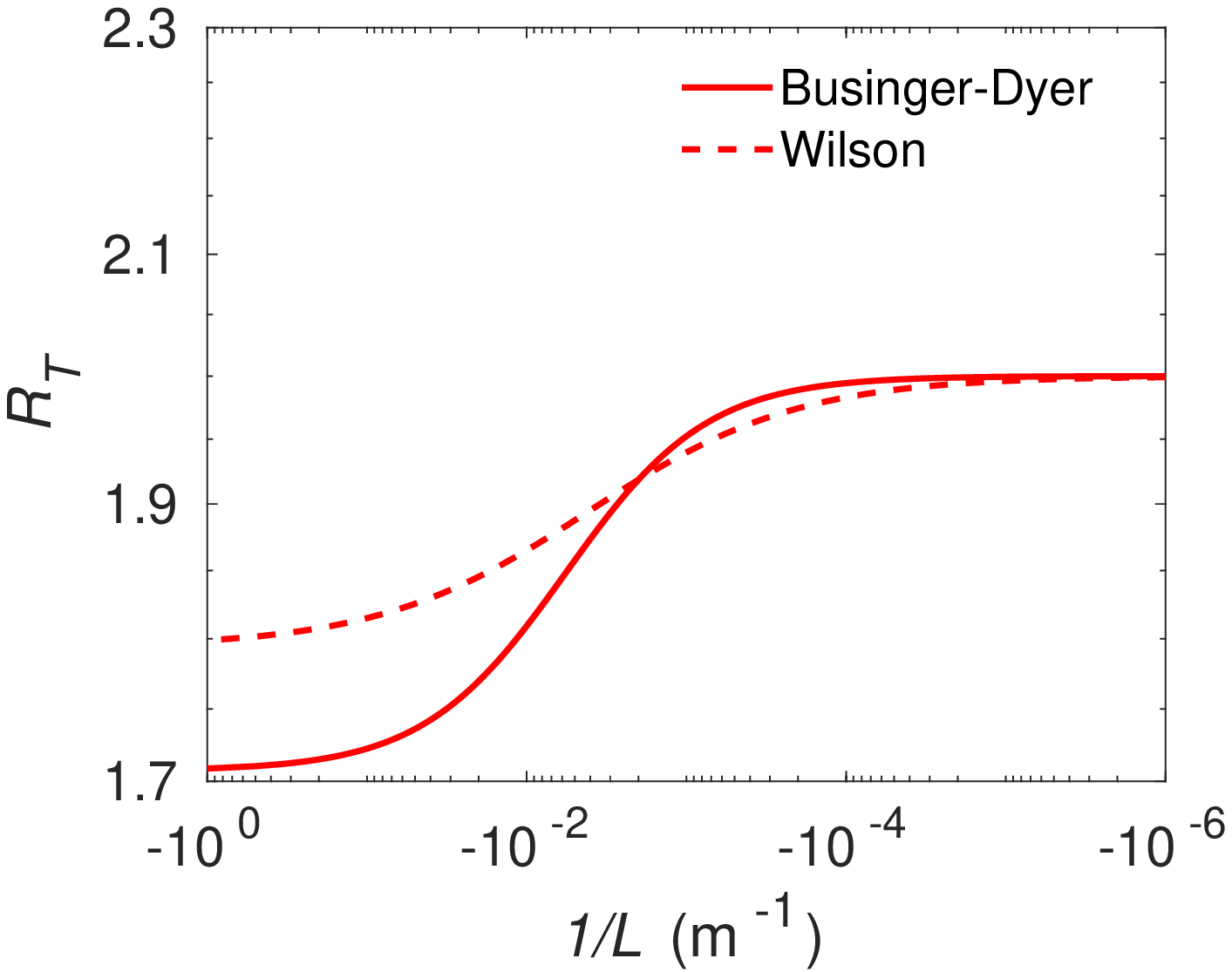}
\includegraphics[width=0.49\textwidth]{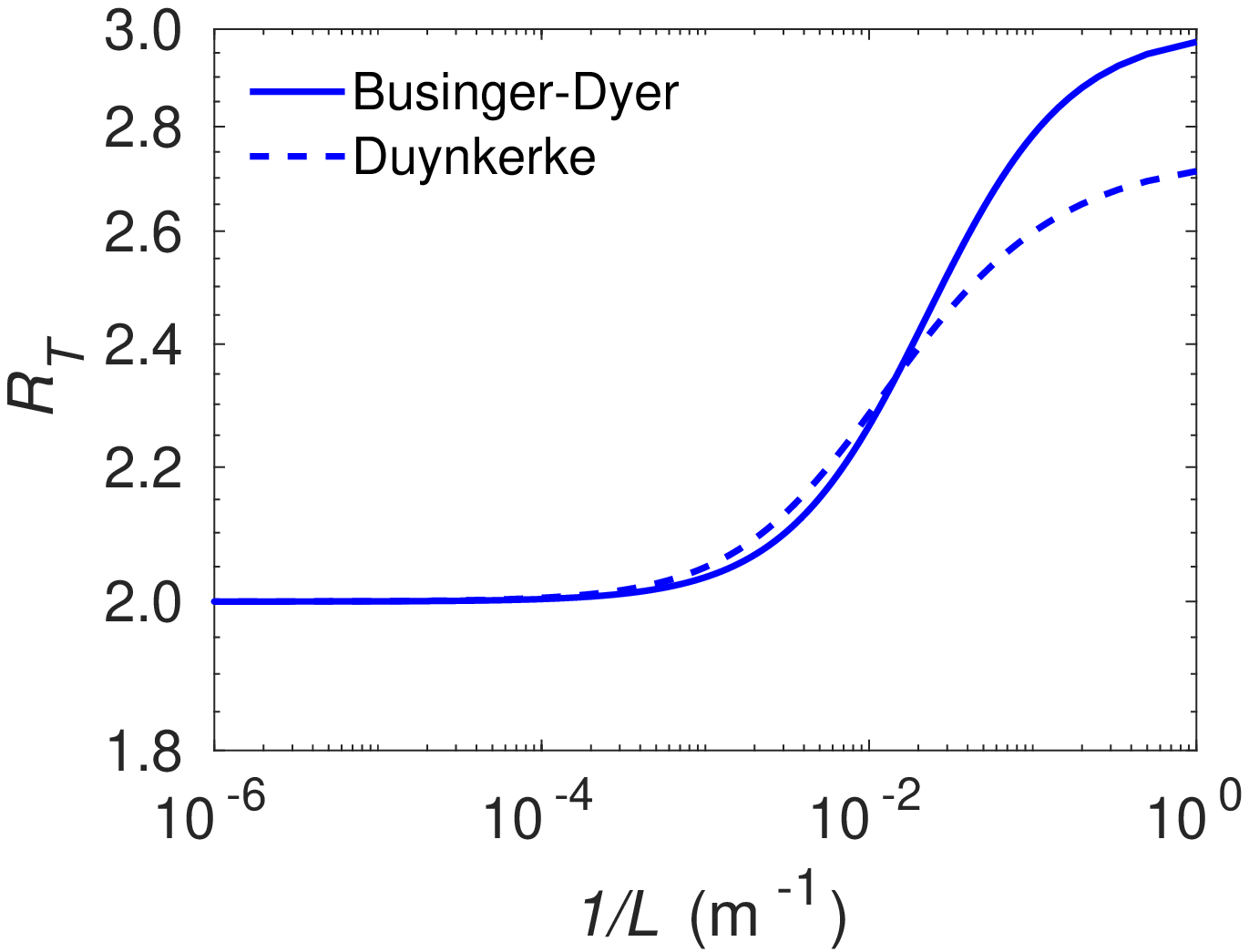}
\caption{Variations of $R_W$ (top panel) and $R_T$ (bottom panel) with respect to inverse Obukhov length ($1/L$). The left and right panels represent unstable and stable conditions, respectively. The legends in these plots correspond to the selected stability correction functions.}
\label{SingleVF}       
\end{figure*}

In Fig.~\ref{SingleVF}, the variations of these ratios with respect to $1/L$ are shown, where several well-known $\psi_m$ and $\psi_h$ functions are utilized in these plots. More details on these functions can be found in Appendix~3. In these plots, the sensor heights are assumed to be at $z_1 = 5$~m, $z_2 = 10$~m, and $z_3 = 20$~m, respectively. For these specific heights, $R_N = 2$. Clearly, for unstable conditions (left panel), both $R_W$ and $R_T$ monotonically decrease with increasing instability. In contrast, for stable conditions (right panels), these ratios show a monotonically increasing trend with increase in stability. For Businger--Dyer functions \citep{Dyer70,Businger71,Dyer74}, it can be readily deduced that both $R_W$ and $R_T$ should approach constant values under very stable conditions, 
\begin{equation}
R_{VS} \rightarrow \frac{\left(z_3 - z_1\right)}{\left(z_2 - z_1\right)}.
\label{RatioVS}
\end{equation} 
For the chosen sensor heights, $R_{VS} = 3$. This asymptotic behaviour is prominently evident in the right panels of Fig.~\ref{SingleVF}. 

In summary, for the selected stability correction functions, $R_W$ and $R_T$ are single-valued functions of $L$. Thus, it should be straightforward to estimate $L$ given measured value of either$R_W$ and $R_T$. In this regard, any suitable root-finding algorithm (e.g., Newton--Raphson approach) can be utilized; we make use of the well-known Levenberg--Marquardt algorithm. Once $L$ is estimated, one can estimate $u_*$ from Eqs.~\ref{dU1} and \ref{dU2}. Since there are two equations and only one unknown, the conventional linear regression approach with ordinary least squares can be employed. Having determined both $L$ and $u_*$, one can then estimate $\overline{w\theta}$ from the definition of Obukhov length. A similar strategy can be followed in conjunction with $R_T$ as input. Of course, in this case, one solves for $\theta_*$ instead of $u_*$, and from the definition of $L$, one deduces $u_*$, and subsequently, $\overline{w\theta}$. 

\section{Limitations of the Proposed Hybrid Approaches}
\label{Limitations}

Before delving into the results, we would like to mention a few  issues that may limit the applications of the proposed hybrid approaches: 

\subsection{Validity of MOST}

Both the hybrid-W and hybrid-T approaches are deeply rooted in MOST. Hence, they are only applicable when and where MOST is applicable. We would like to remind the readers that MOST is strictly valid in a horizontally homogeneous surface layer. In the surface layer (aka constant flux layer), the turbulent fluxes are assumed to be invariant with height. Thus, all the sensor heights (i.e., $z_1$ , $z_2$, $z_3$) should be within the surface layer to avoid violation of MOST. For strongly stratified conditions, the surface layer may be only a few metres deep; the proposed hybrid approaches should be avoided under that scenario. 

\subsection{Monotonicity of Input Mean Profiles}

The hybrid-W approach implicitly assumes that wind speeds monotonically increase with height. Similarly, in the case of the hybrid-T approach, the potential temperature is expected to monotonically increase (decrease) with height for stable (unstable) conditions. If such monotonic conditions are not met, the proposed approaches should not be used. 

\subsection{Similarity of Footprints}

The footprints for scalars and fluxes should be similar in order to estimate fluxes accurately via MOST; over homogeneous surface conditions, this restriction is not that important. However, for heterogeneous cases, the mismatch of footprints could pose a serious limitation. Of course, any application of MOST for these cases will also be questionable. 

\subsection{Multi-valued Functions}

\begin{figure*}[ht]
\centering
\includegraphics[width=0.49\textwidth]{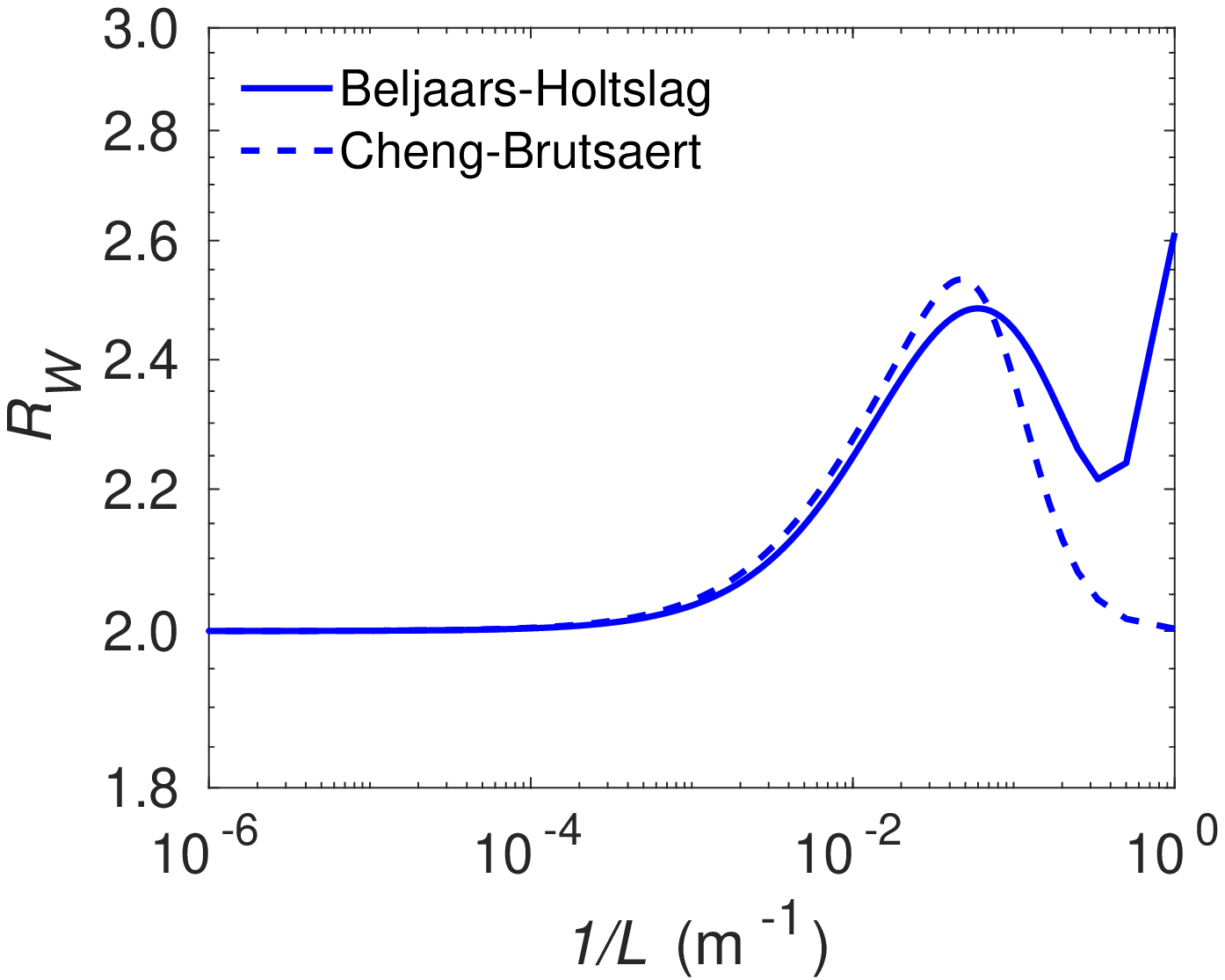}
\includegraphics[width=0.49\textwidth]{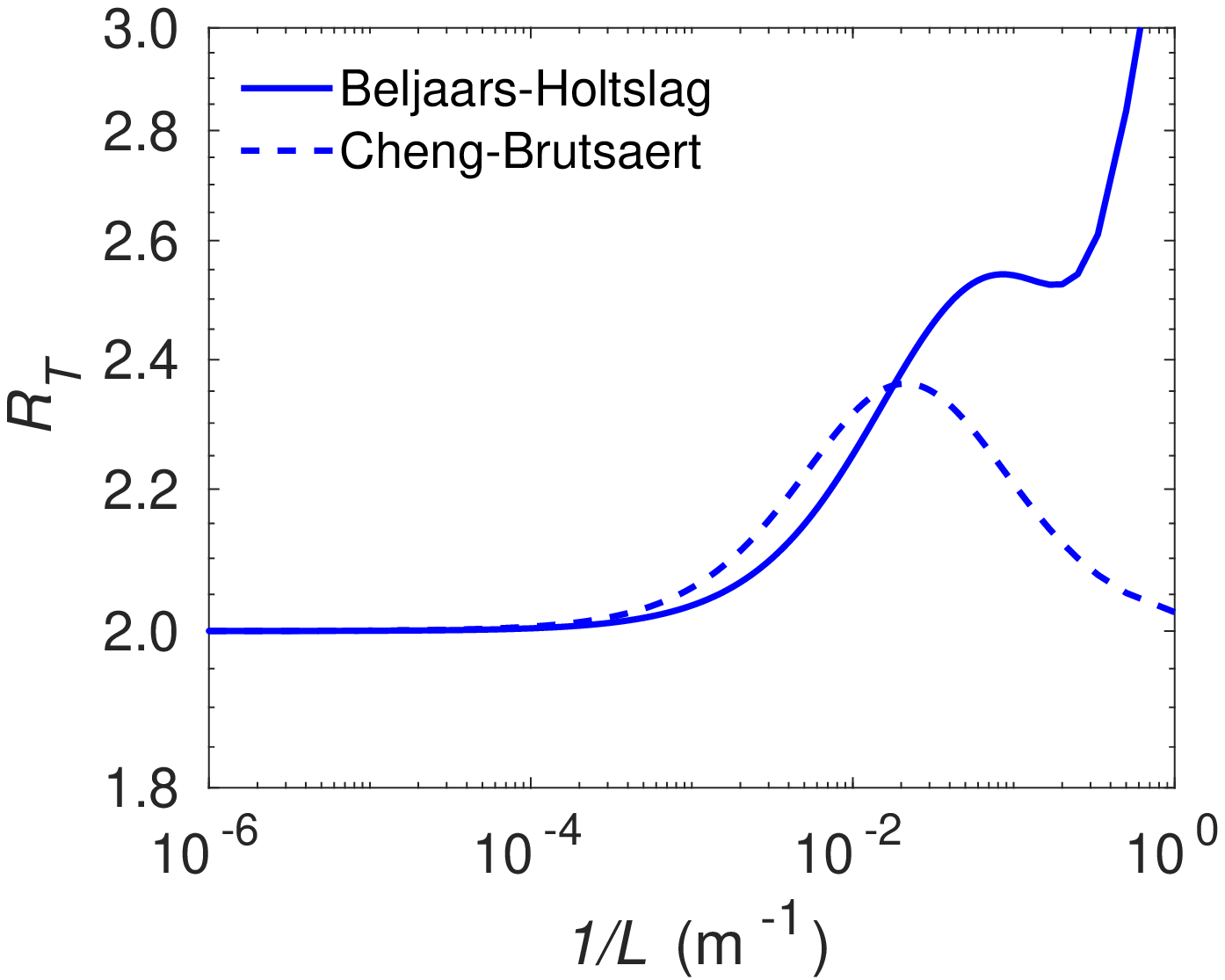}
\caption{Variations of $R_W$ (left panel) and $R_T$ (right panel) with respect to inverse Obukhov length ($1/L$). The legends in these plots correspond to the selected stability correction functions.}
\label{MultiVF}       
\end{figure*}

In Sect.~3, we have shown that $R_W$ and $R_T$ variables are single-valued functions of $L$ for a specific set of widely used stability correction functions. However, there are exceptions. In Fig.~\ref{MultiVF}, we compute the same ratios using stability correction formulations proposed by Beljaars and Holtslag~\cite{Beljaars91} and Cheng and Brutsaert~\cite{Cheng05} for stably stratified conditions (see Appendix~3 for details). Clearly, the resultant functions are multi-valued; in other words, given $R_W$ or $R_T$ , it is not possible to estimate unique values of $L$. As a consequence, our proposed hybrid approach should not be used in conjunction with these specific stability correction functions. 

\subsection{Turbulent Prandtl Number}

In the MOST relation for the potential temperature profile, Eq.~\ref{MOST_T}, we implicitly assume that the turbulent Prandtl number ($Pr_T$) is equal to one. Since the estimation of $L$ only depends on the ratio $R_T$, this assumption is not relevant. However, its influence on the estimations of $\theta_*$ and $u_*$ via hybrid-T approach cannot be disregarded. Note that the hybrid-W approach does not involve any information about $Pr_T$. 

\subsection{Effects of Moisture}

Throughout this paper, we only considered dry atmospheric conditions in the surface layer. It is, however, straightforward to extend the hybrid approaches for moist conditions (e.g., offshore environments). In these cases, one must utilize virtual kinematic heat flux and the virtual potential temperature in the definition of Obukhov length ($L$) and in Eq.~\ref{MOST_T}. The stability parameter ($z/L$) can even be partitioned to account for sensible heat flux and latent heat flux separately. For further details, see Barthelmie et al.~\cite{Barthelmie10} and the references therein. 

\section{Inter-comparison of Different Flux-Estimation Approaches}
\label{Comparison}

In order to compare the proposed hybrid approaches against the traditional ones, we perform Monte-Carlo-type numerical experiments with the following steps:  

\begin{enumerate}[(i)]
\item To encompass a wide-range of stability conditions, we assume $u_* \in \left[0.1~2\right]$ m~s$^{-1}$ and $\theta_* \in \left[-1~0.2\right]$ K. From these sets, we randomly (with uniform probability) select a $\left(u_*, \theta_*\right)$ pair. 
\item Furthermore, we assume $z_0 = z_{0 T} = $ 0.1 m and $\Theta_s = \Theta_0 =$ 300 K. 
\item Using these selected inputs, we first estimate $L$, and then in turn, predict $U\left(z\right)$ and $\Theta\left(z\right)$ via Eqs.~\ref{MOST_U} and \ref{MOST_T} in conjunctions with the Businger--Dyer stability correction functions [i.e., Eqs.~\ref{psiBusinger}a--c].
\item In `noise-free input data' cases, we skip this specific step. Otherwise, we add random noise on $U(z)$ and $\Theta(z)$ profiles. More details on the characteristics of additive noise are provided later.  
\item If the estimated $z/|L| < 1$ and mean wind speed $> 1$ m~s$^{-1}$, then, we proceed to the following step. Otherwise, we discard the selected $\left(u_*, \theta_*\right)$ pair and go back to the first step. In the `noisy input data' cases, we enforce a few more additional exclusion criteria which will be discussed later.  
\item Next, we attempt to do the following inverse computation: given the predicted mean wind-speed and/or temperature profiles, can we accurately estimate the surface fluxes? In hybrid-W (hybrid-T) approach, we estimate the surface fluxes by only using wind-speed (potential temperature) data from $z$ = 5, 10, and 20 m.
\item In order to have a direct comparison, we also estimate fluxes using the traditional gradient and profile methods (Appendix~1). In this case, both wind and temperature data from the lowest two levels are utilized. 
\item  For all the flux-estimation approaches, we quantify the relative errors in the estimations of $u_*$ and $\theta_*$.
\item We repeat all the previous steps until we get $10^5$ admissible samples for all the scenarios.   
\end{enumerate}

\subsection{Noise-free Input Data}

The relative errors for the estimations of $u_*$ and $\theta_*$ are reported in Table~\ref{T1}. These errors are computed as follows,
\begin{equation}
RE = \frac{\chi_{\mbox{est}}-\chi_{\mbox{true}}}{\chi_{\mbox{true}}}\times 100
\end{equation}
where $\chi$ is either $u_*$ or $\theta_*$. In addition to minimum and maximum values, several percentiles (based on $10^5$ samples for each case) are reported in Table~\ref{T1}. 

\begin{table}[h]
\caption{Relative errors (\%) in the estimations of $u_*$ and $\theta_*$}
\label{T1}
\begin{tabular}{lccccccc}
\hline\noalign{\smallskip}
 & min & $p_1$ & $p_{25}$ & $p_{50}$ & $p_{75}$ & $p_{99}$ & max\\   
\noalign{\smallskip}\hline\noalign{\smallskip}
      & \multicolumn{7}{c}{Estimation of ${u_*}$}\\
Hybrid-W & $-$2.1$\times10^{-3}$ & 0 & 0 & 0 & 0 & 0 & 3.0$\times10^{-4}$ \\
Hybrid-T & $-$90.1 & 0 & 0 & 0 & 0 & 0 & 9.3$\times10^{-4}$ \\
Gradient & 4.0 & 4.0 & 4.0 & 4.0 & 4.1 & 4.5 & 4.5 \\
Profile & 0 & 0 & 0 & 0 & 0 & 0 & 0 \\
\\
      & \multicolumn{7}{c}{Estimation of ${\theta_*}$}\\
Hybrid-W & 0 & 0 & 0 & 0 & 0 & 0 & 5100 \\
Hybrid-T & 0 & 0 & 0 & 0 & 0 & 0 & 0 \\
Gradient & 0 & 4.0 & 4.0 & 4.1 & 4.4 & 5.1 & 8.3 \\
Profile & 0 & 0 & 0 & 0 & 0 & 0 & 0 \\
\noalign{\smallskip}\hline
\end{tabular}
\end{table}

Clearly, for both $u_*$ and $\theta_*$, the performance of the traditional profile method is the best among all the approaches as it leads to null errors. In contrast, the traditional gradient method seems to suffer from a systematic error of $O(4\%)$. This error stems from finite-difference approximations, as discussed by Arya~\cite{Arya91}.   

For both hybrid approaches, the relative errors equal zero for percentiles ranging from 1 to 99. In the case of hybrid-W approach, negligible errors can occur in the estimation of $u_*$ due to round off errors during the optimization process. In the case of $\theta_*$, only 17 samples (out of $10^5$) exceeded errors $>$ 1\%. Most of these cases had true $\theta_*$ values close to zero and the division by a small number led to very large relative errors. The performance of the hybrid-T approach was perfect for the estimation of $\theta_*$. In the case of $u_*$ estimation, 16 samples (out of $10^5$) exceeded absolute relative error of 1\%. In summary, for the noise-free cases, the overall performance of the proposed hybrid approaches is almost at par with the traditional profile method. In the following sub-section, we investigate if this conclusion holds in the presence of random errors in input mean profiles.

\subsection{Noisy Input Data}

We conduct uncertainty propagation experiments to quantify if and how the errors in the input profiles are amplified during various flux estimations. We first add different amounts of noise to the profiles as follows, 

\begin{subequations}
\begin{equation}
\widetilde{U} = U + \eta_U,
\end{equation}
\begin{equation}
\widetilde{\Theta} = \Theta + \eta_\Theta.
\end{equation}
\end{subequations}
The noise terms ($\eta_U$ and $\eta_\Theta$) are generated from a multivariate Gaussian distribution with zero mean and the following covariance matrix, 
\begin{equation}
\Sigma = \sigma^2\begin{bmatrix} 
1 & \rho & \rho \\ 
\rho & 1 & \rho \\
\rho & \rho & 1
\end{bmatrix},
\end{equation}
where, $\sigma^2$ is the variance of the noise term. Since we are only concerned with three levels of observations, $\Sigma$ is a $3\times3$ matrix. The variable $\rho$ captures the correlation of noise between different levels. Such a correlated noise situation is possible when a single instrument (e.g., a lidar) is used to measure wind speeds (or temperature) at different heights.  

\begin{table}[h]
\caption{Different scenarios for the noise terms}
\label{T2}
\begin{tabular}{lcccc}
\hline\noalign{\smallskip}
      & \multicolumn{2}{c}{$\eta_U$} & \multicolumn{2}{c}{$\eta_\Theta$} \\
Scenario & $\sigma$ (m s$^{-1}$) & $\rho$  & $\sigma$ (K) & $\rho$ \\
\noalign{\smallskip}\hline\noalign{\smallskip}
1      & 0.01     & 0.9  & -     & - \\
2      & 0.01     & 0.5  & -     & - \\
3      & 0.05     & 0.9  & -     & - \\
4      & 0.05     & 0.5  & -     & - \\
5      & 0.05     & 0.5  & 0.01  & 0.9 \\
6      & 0.05     & 0.5  & 0.05  & 0.5 \\
\noalign{\smallskip}\hline
\end{tabular}
\end{table}

We consider several noise scenarios which are listed in Table~\ref{T2}. Specifically, we consider two noise levels (with appropriate units): 0.01 (low) and 0.05 (high). In addition, two values of $\rho$ are considered: 0.9 (high) and 0.5 (low). Since the hybrid-W approach only requires wind-speed data, please note that the scenarios 4, 5, and 6 are all the same for this approach. 

\begin{figure*}[ht]
\centering
\includegraphics[height=1.75in]{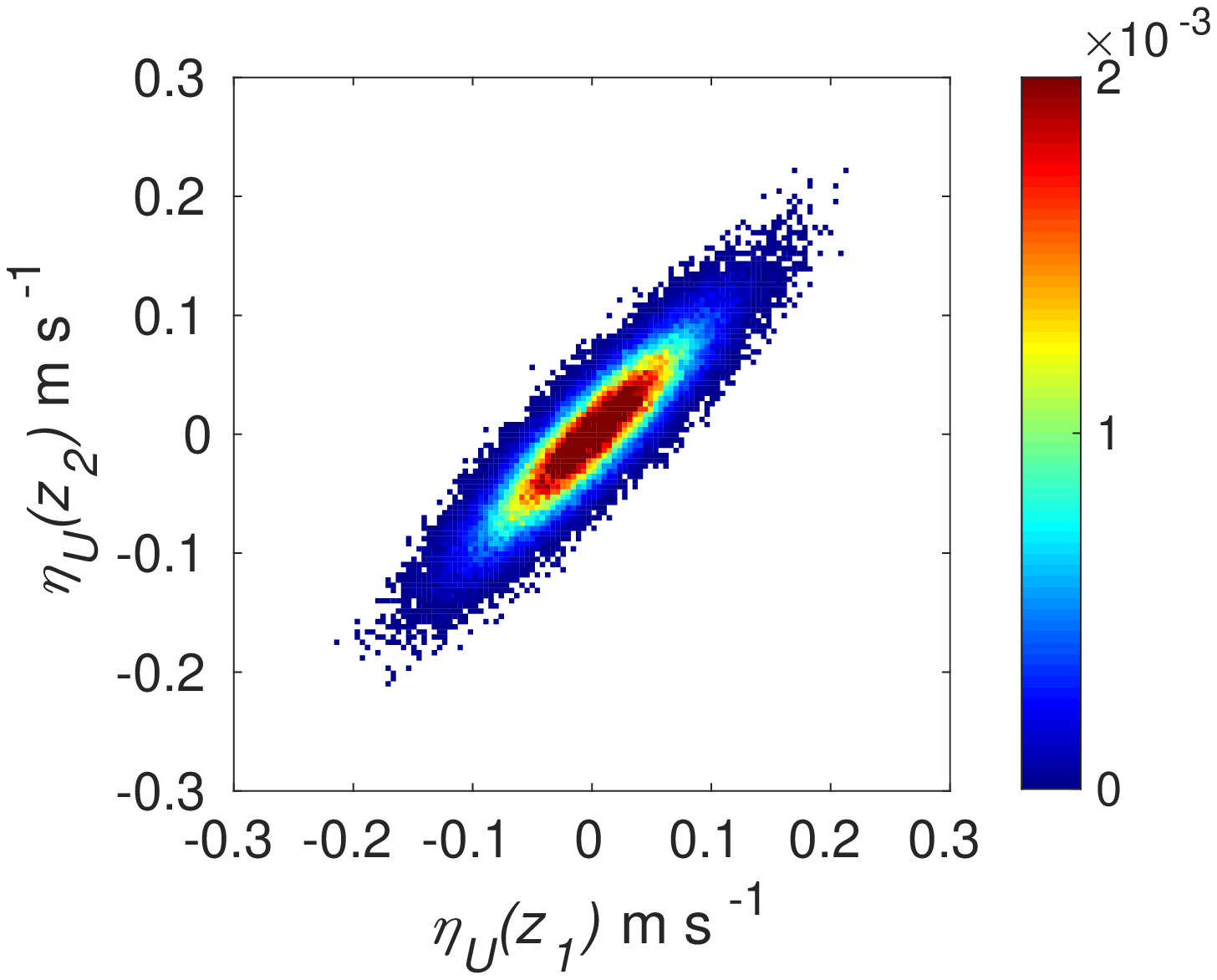}
\hspace{0.1in}
\includegraphics[height=1.75in]{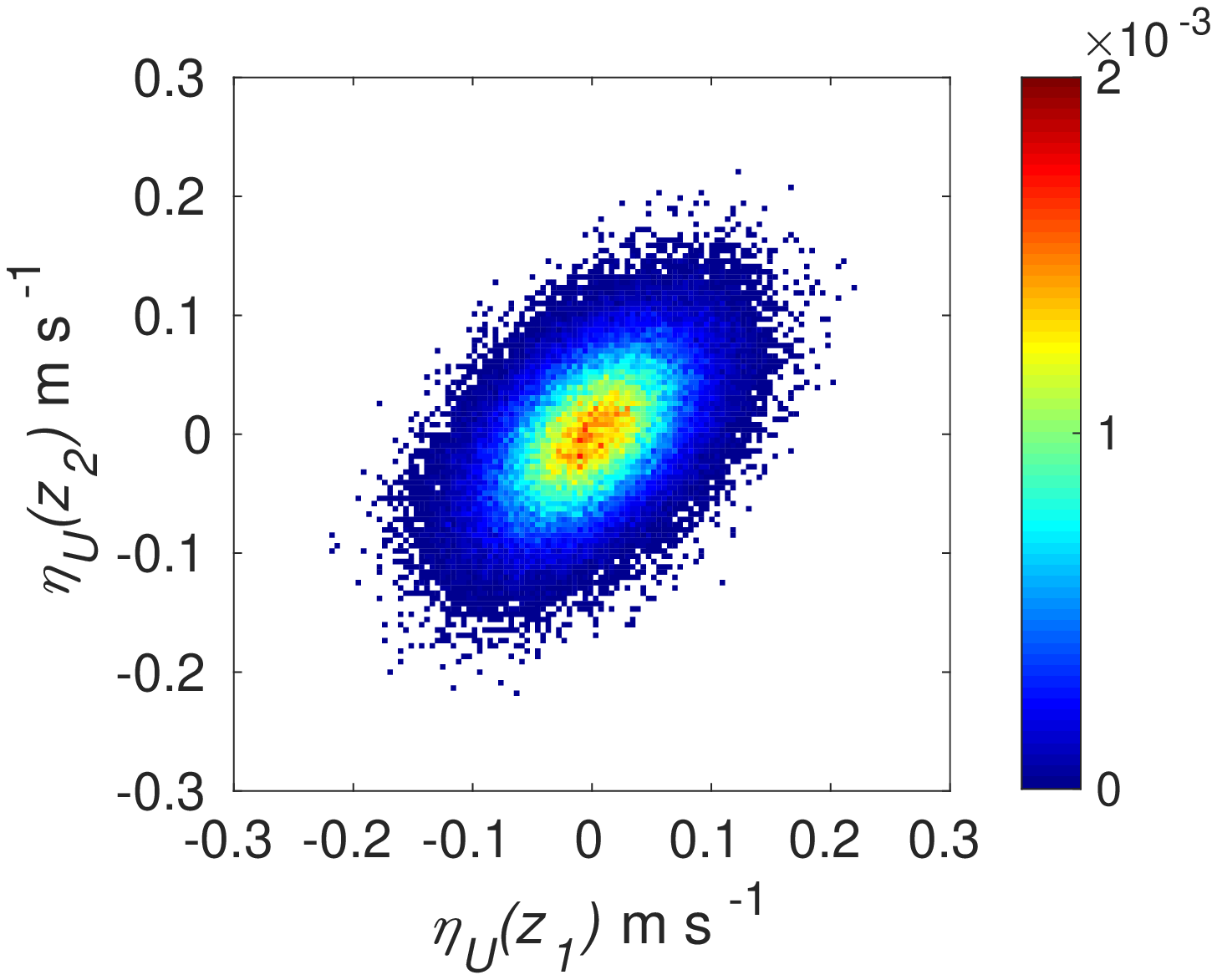}
\caption{Bi-variate probability density functions for scenarios 3 (left panel) and 4 (right panel), respectively. For both the scenarios, $\sigma = $ 0.05 m s$^{-1}$. However, $\rho$ changes from 0.9 in scenario 3 to 0.5 in scenario 4. Here $z_1$ and $z_2$ denote two different sensor heights.}
\label{Noise}       
\end{figure*}

Illustrative noise values ($\eta_U$) are shown in Fig.~\ref{Noise}. Clearly, for $\sigma = 0.05$ m~s$^{-1}$, the noise terms can reach up to $\pm 0.2$ m~s$^{-1}$. Large amount of additive random noise can distort the $U(z)$ and $\Theta(z)$ profiles significantly and can even make them physically unrealistic. To avoid such undesirable situations, we implemented certain exclusion criteria in addition to the ones discussed in the previous sub-section [i.e., step (v)]. If the noisy $U(z)$ and $\Theta(z)$ profiles are not monotonic, we exclude that particular case. If the resultant $R_W$ and $R_T$ values are outside their acceptable ranges (i.e., $1.8 < R_W < 3$ and $1.7 < R_T < 3$; see Fig.~\ref{SingleVF}), those cases are also excluded. 

\begin{figure*}[ht]
\centering
\includegraphics[height=1.75in]{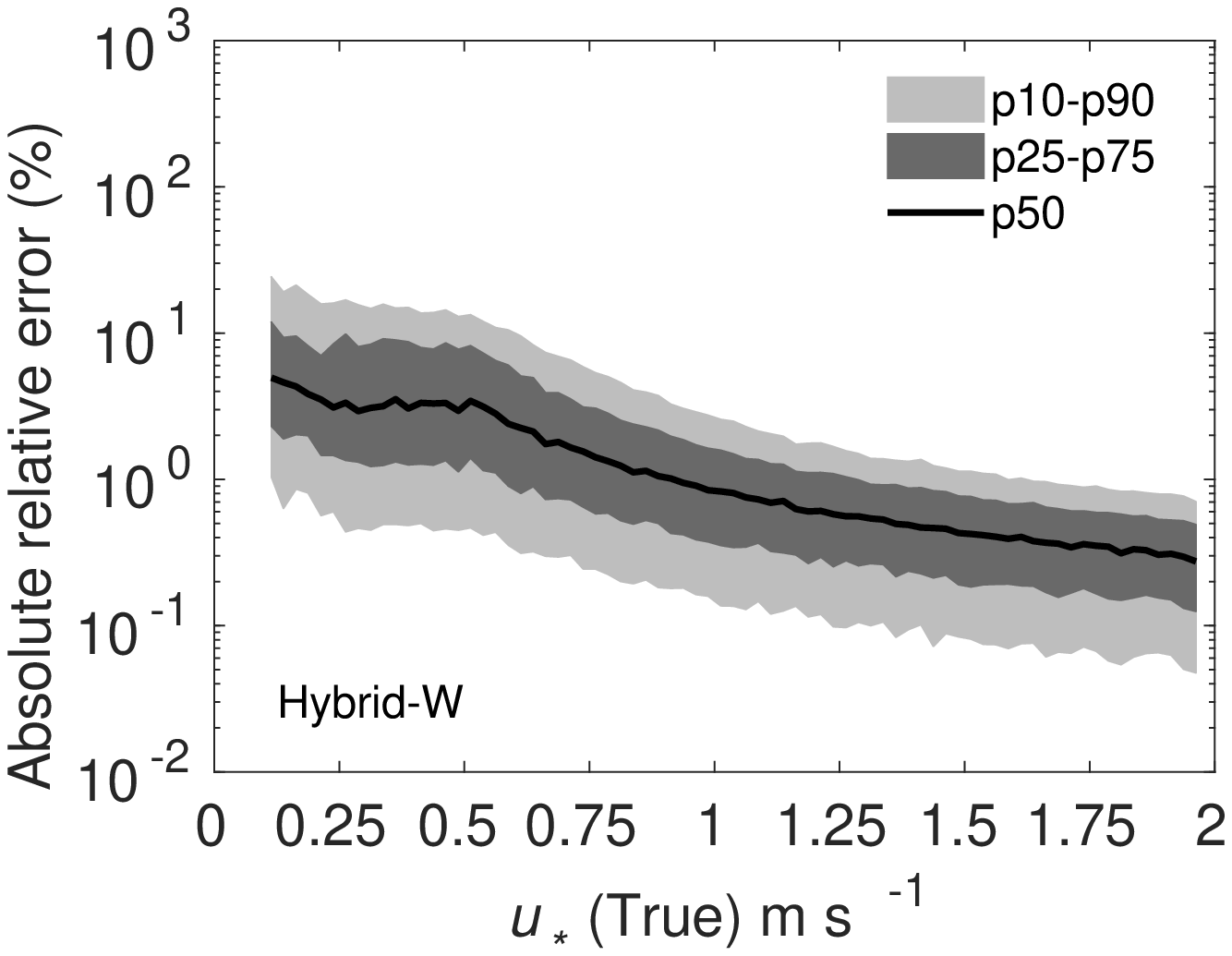}
\includegraphics[height=1.75in]{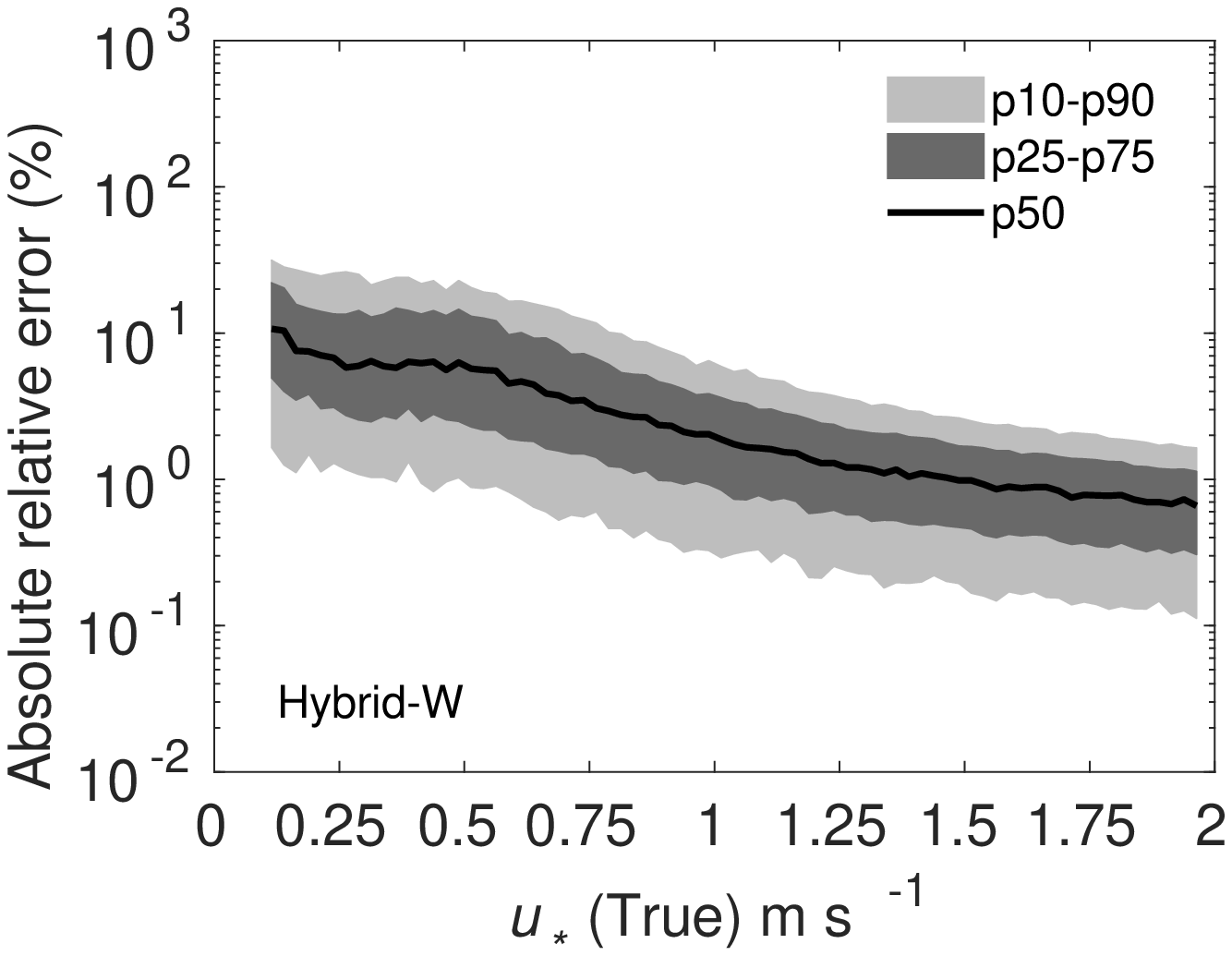}\\
\vspace{0.1in}
\includegraphics[height=1.75in]{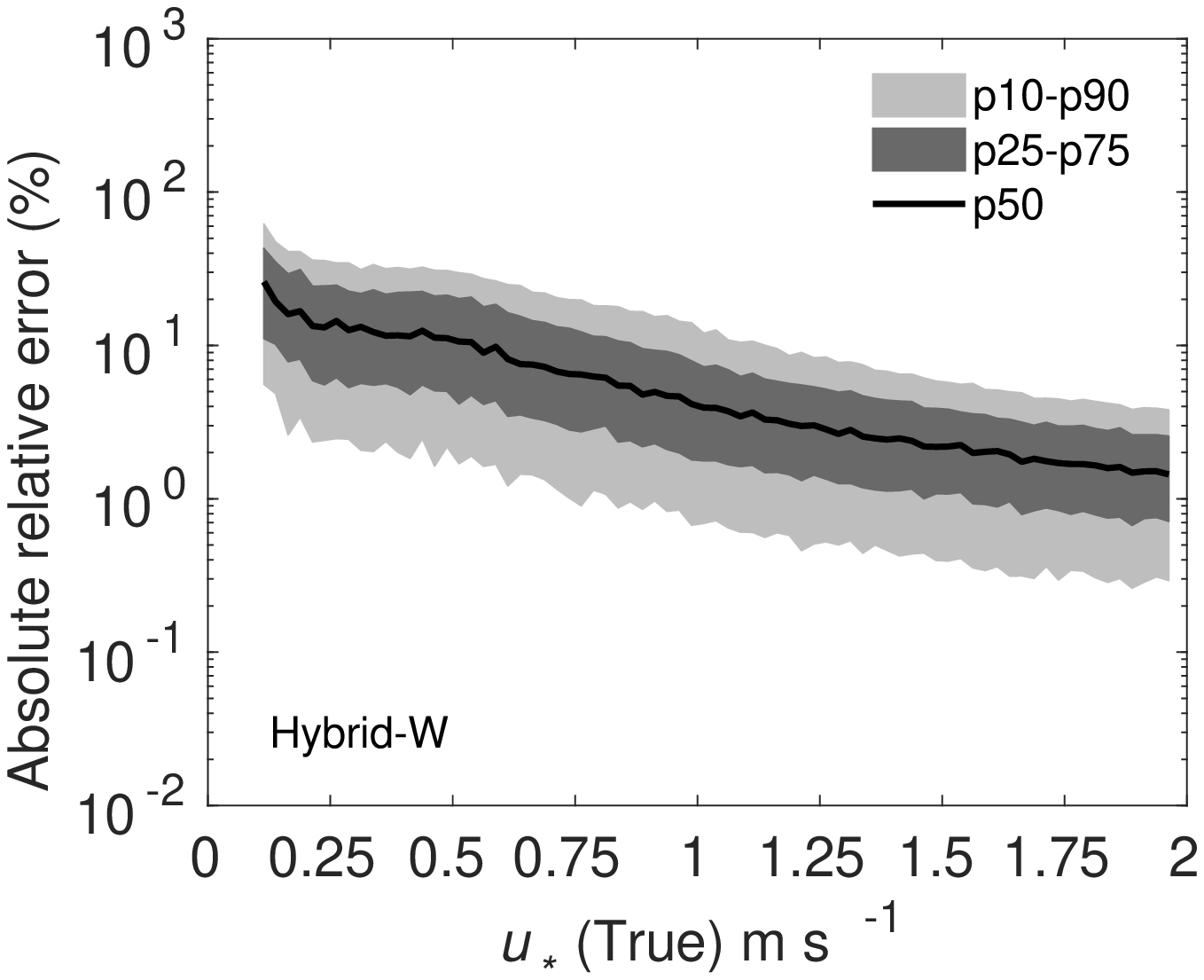}
\includegraphics[height=1.75in]{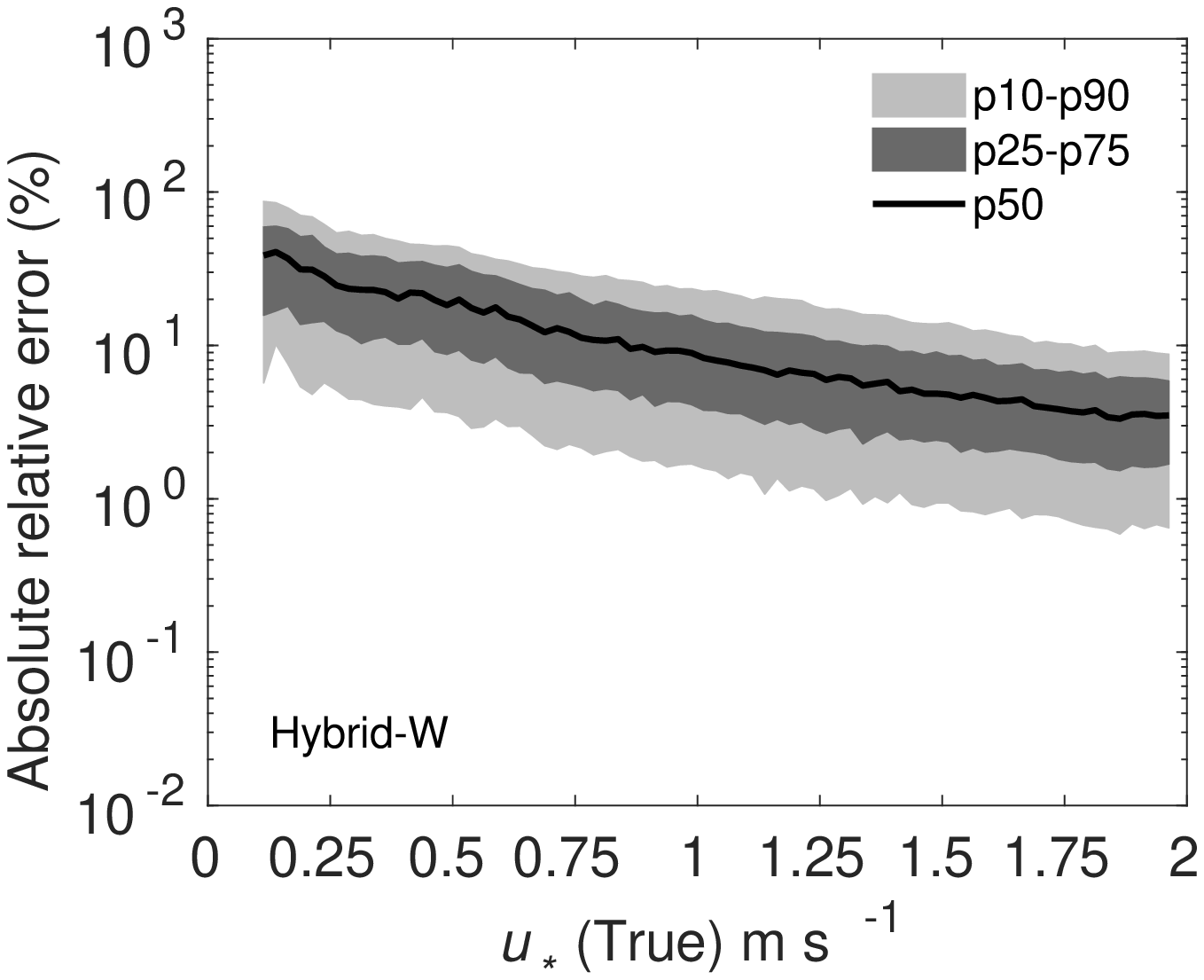}
\caption{Absolute relative errors in the estimation of $u_*$ for four scenarios: 1 (top-left panel), 2 (top-right panel), 3 (bottom-left panel), and 4 (bottom-right panel). Flux-estimation approach: hybrid-W.}
\label{HyW_ustar}       
\end{figure*}

\begin{figure*}[ht]
\centering
\includegraphics[width=0.49\textwidth]{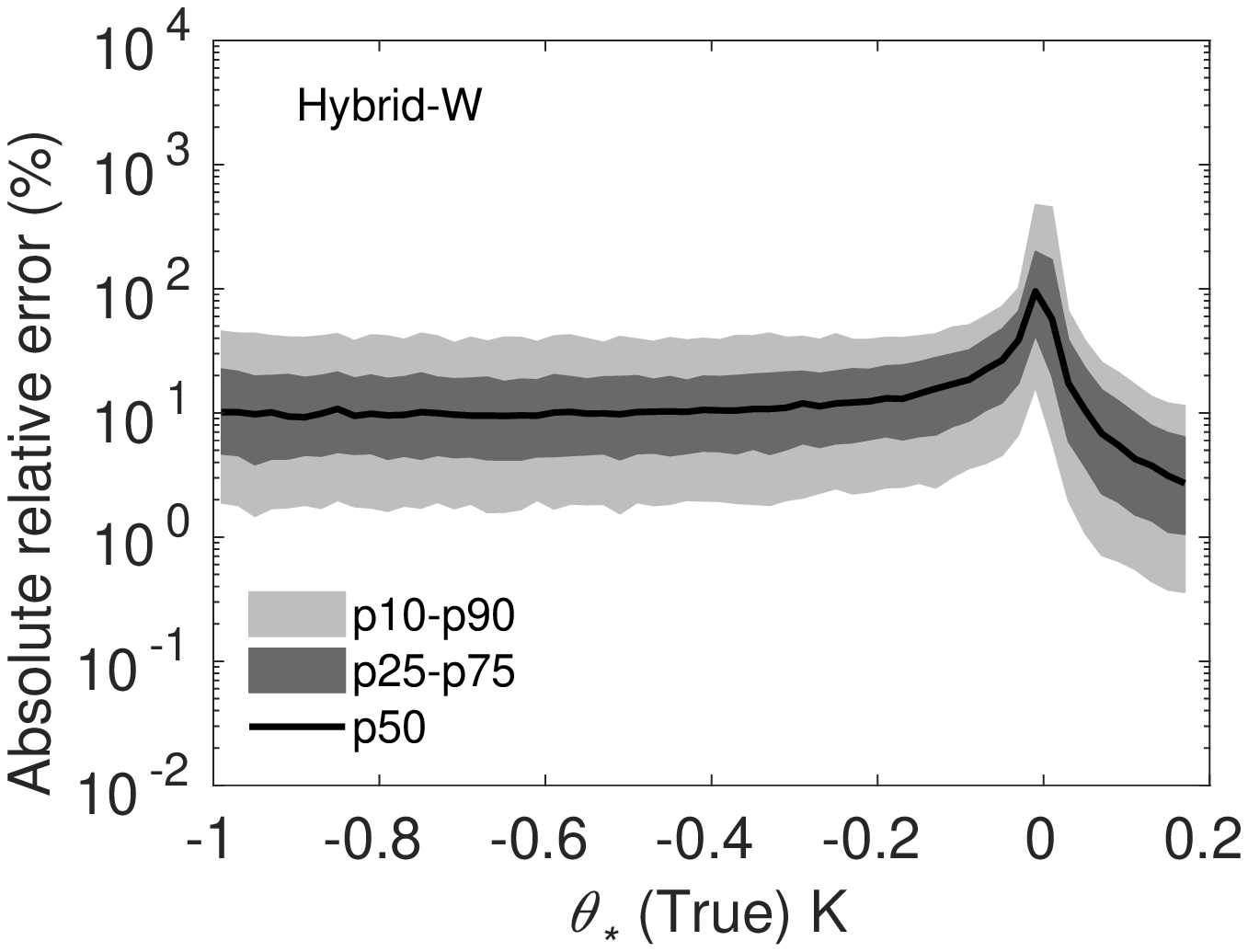}
\includegraphics[width=0.49\textwidth]{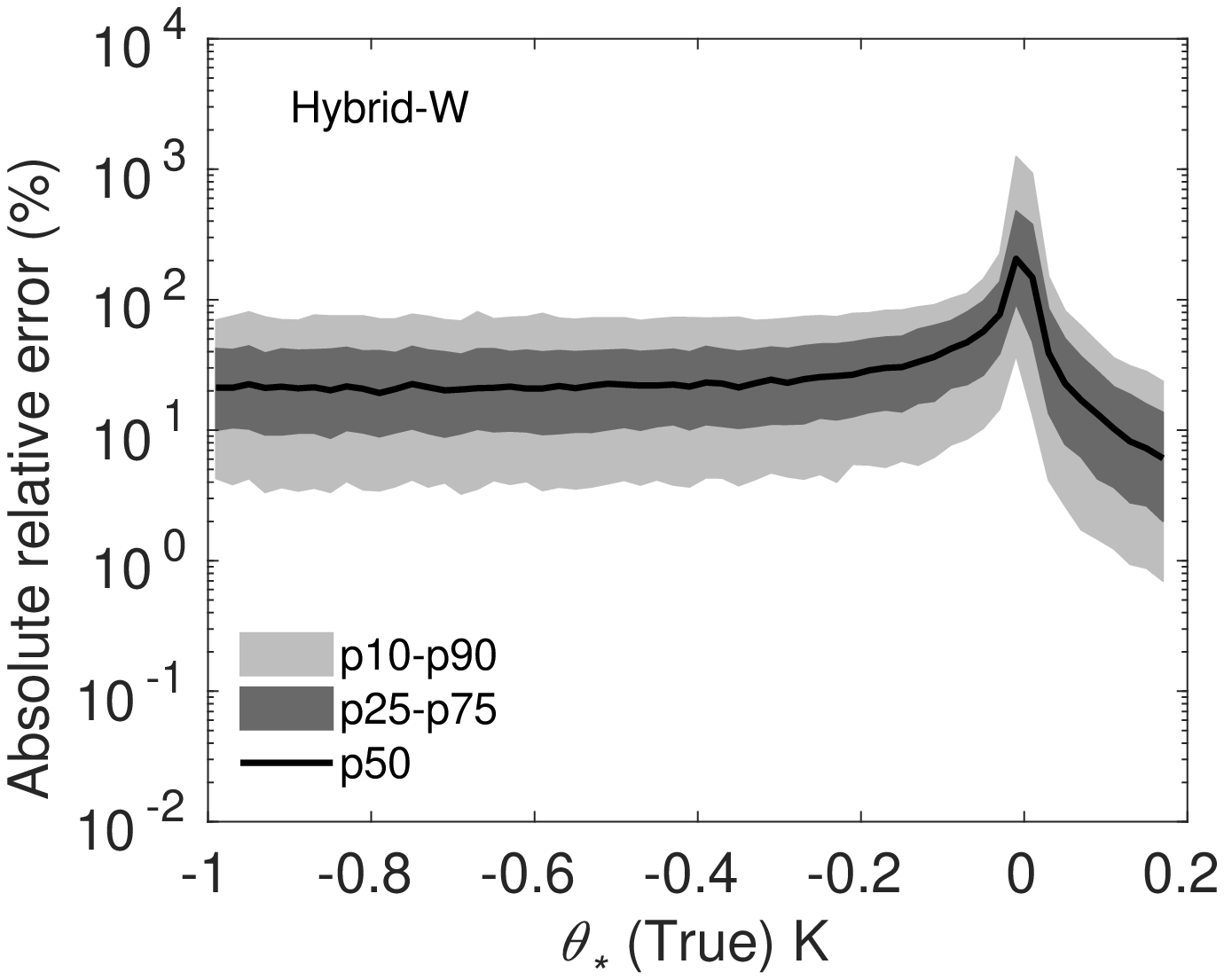}\\
\vspace{0.1in}
\includegraphics[width=0.49\textwidth]{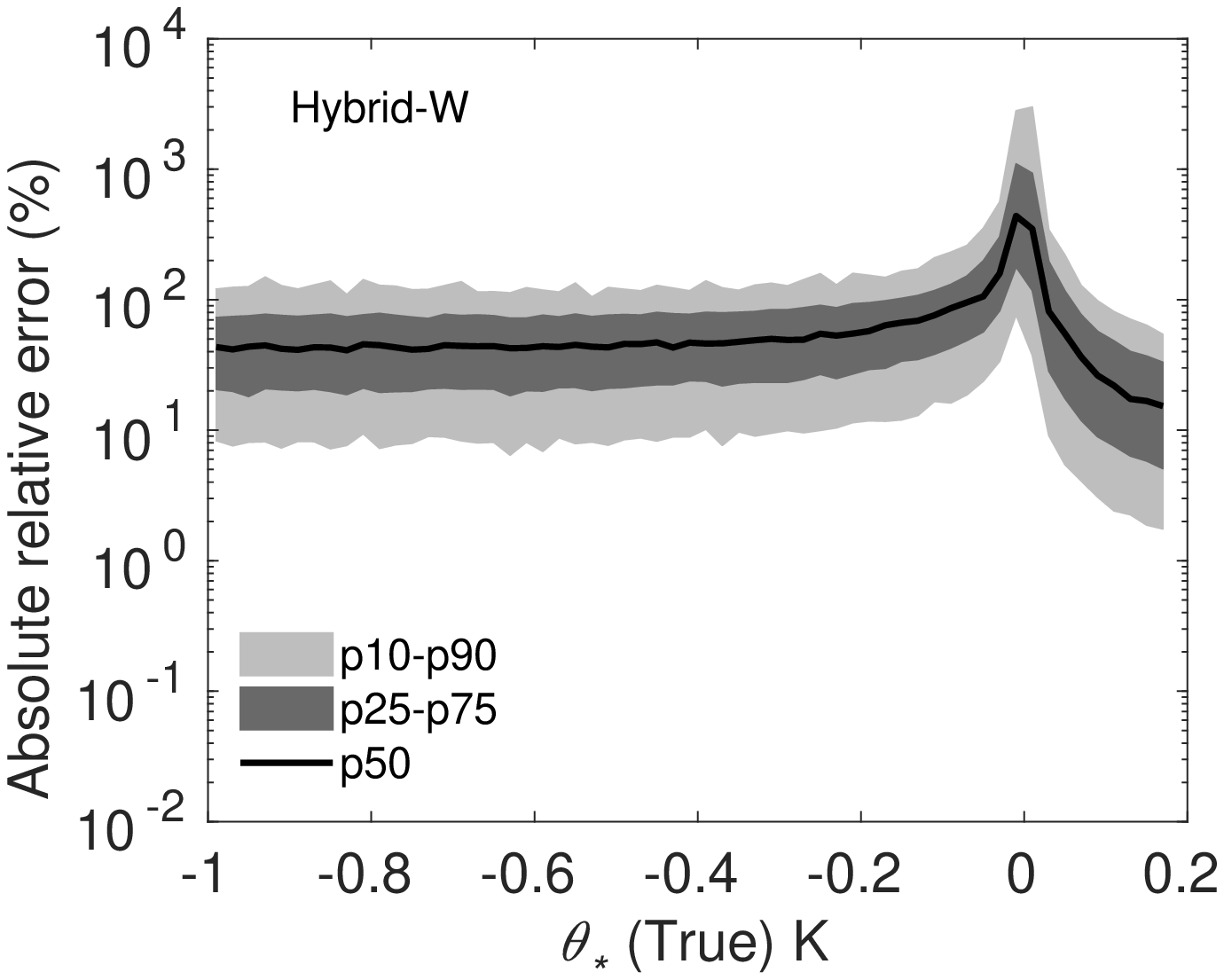}
\includegraphics[width=0.49\textwidth]{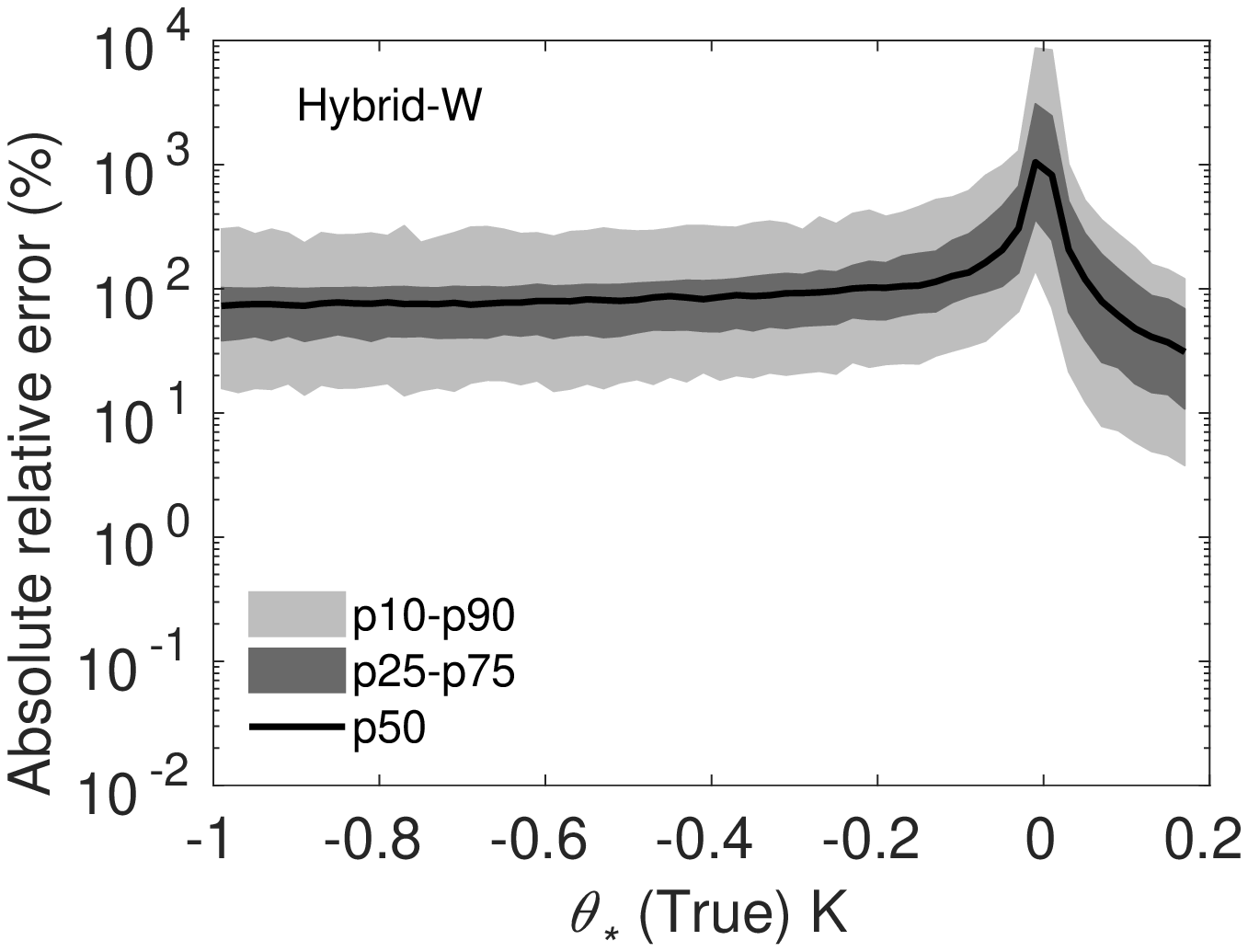}
\caption{Absolute relative errors in the estimation of $\theta_*$ for four scenarios: 1 (top-left panel), 2 (top-right panel), 3 (bottom-left panel), and 4 (bottom-right panel). Flux-estimation approach: hybrid-W.}
\label{HyW_tstar}       
\end{figure*}

\begin{figure*}[ht]
\centering
\includegraphics[width=0.49\textwidth]{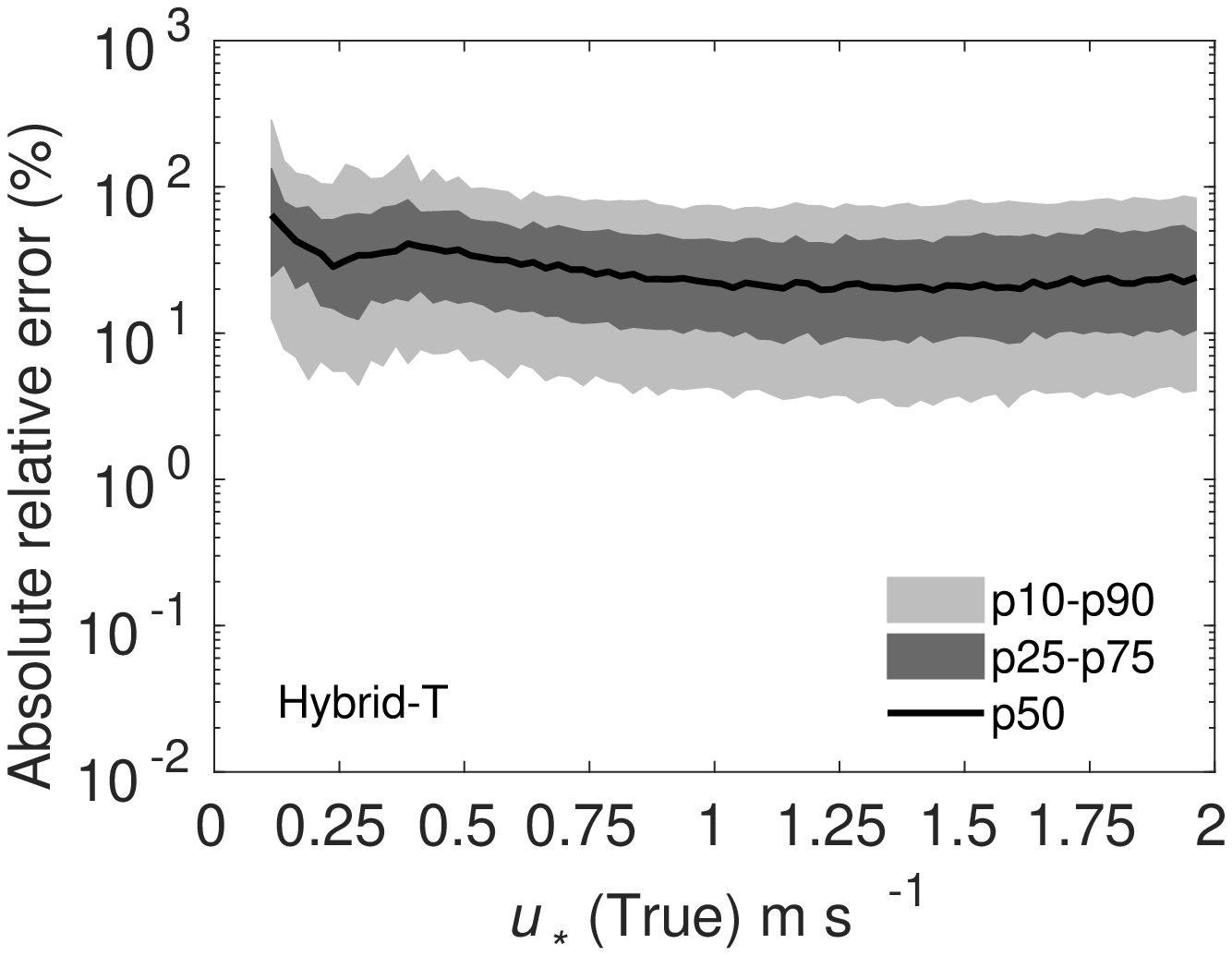}
\includegraphics[width=0.49\textwidth]{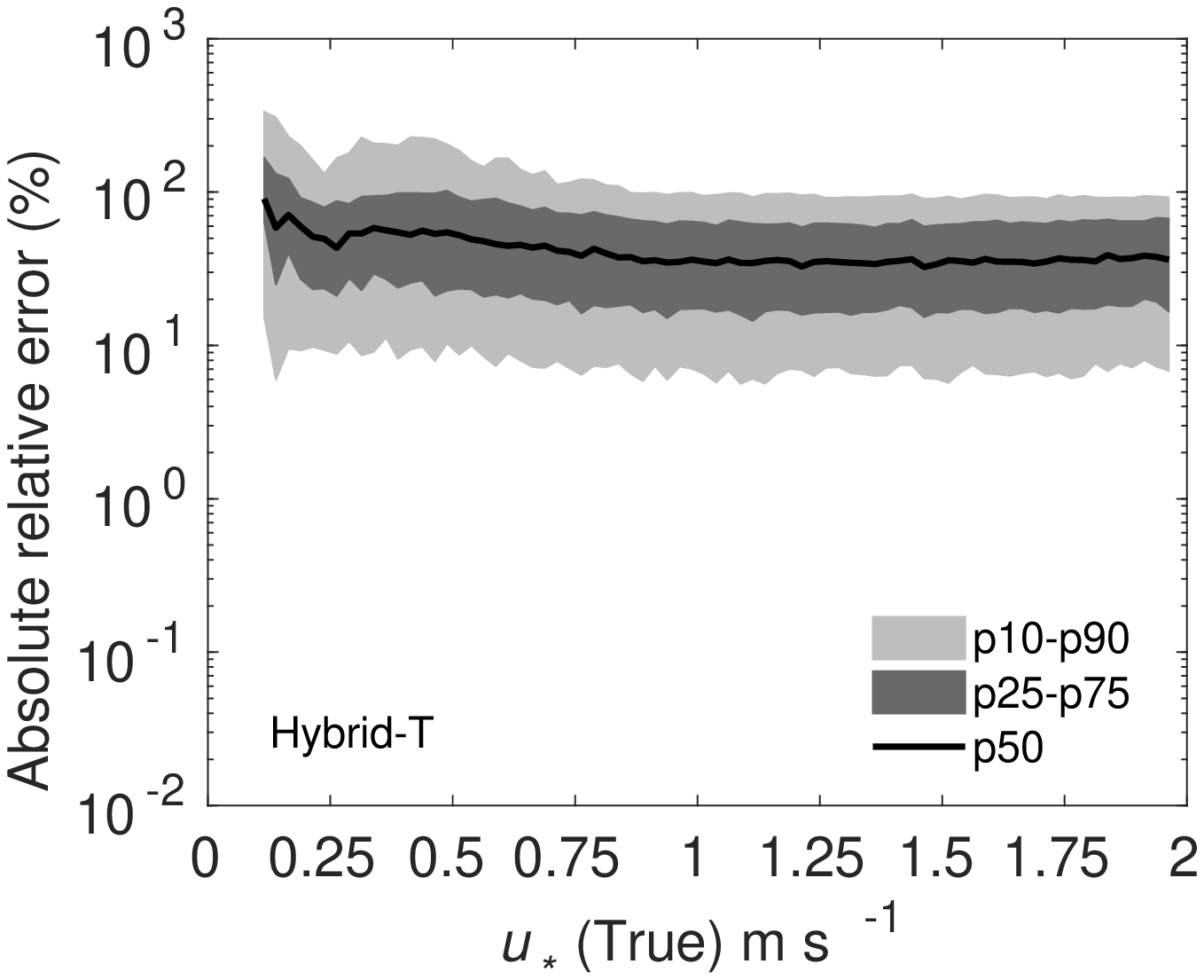}\\
\vspace{0.1in}
\includegraphics[width=0.49\textwidth]{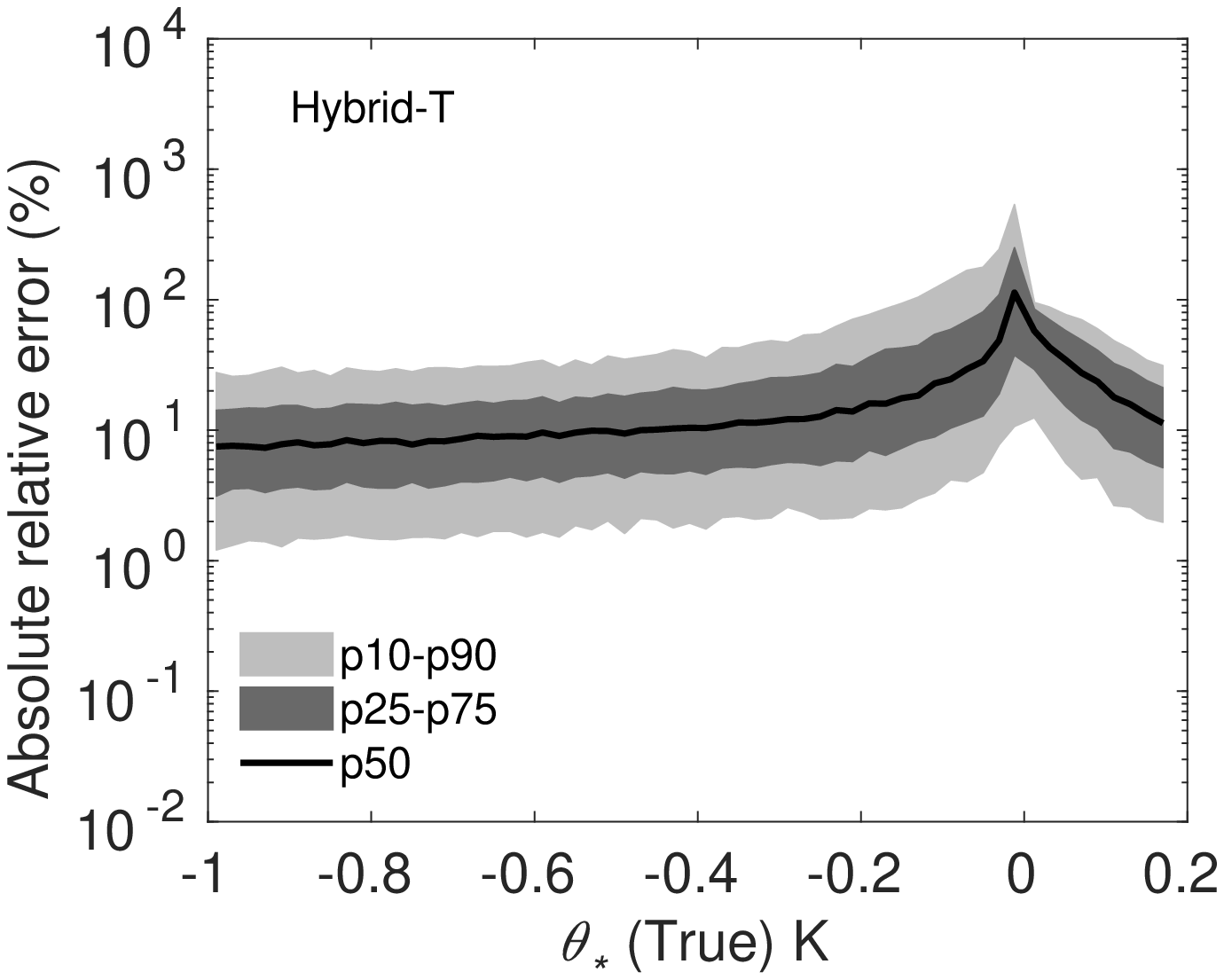}
\includegraphics[width=0.49\textwidth]{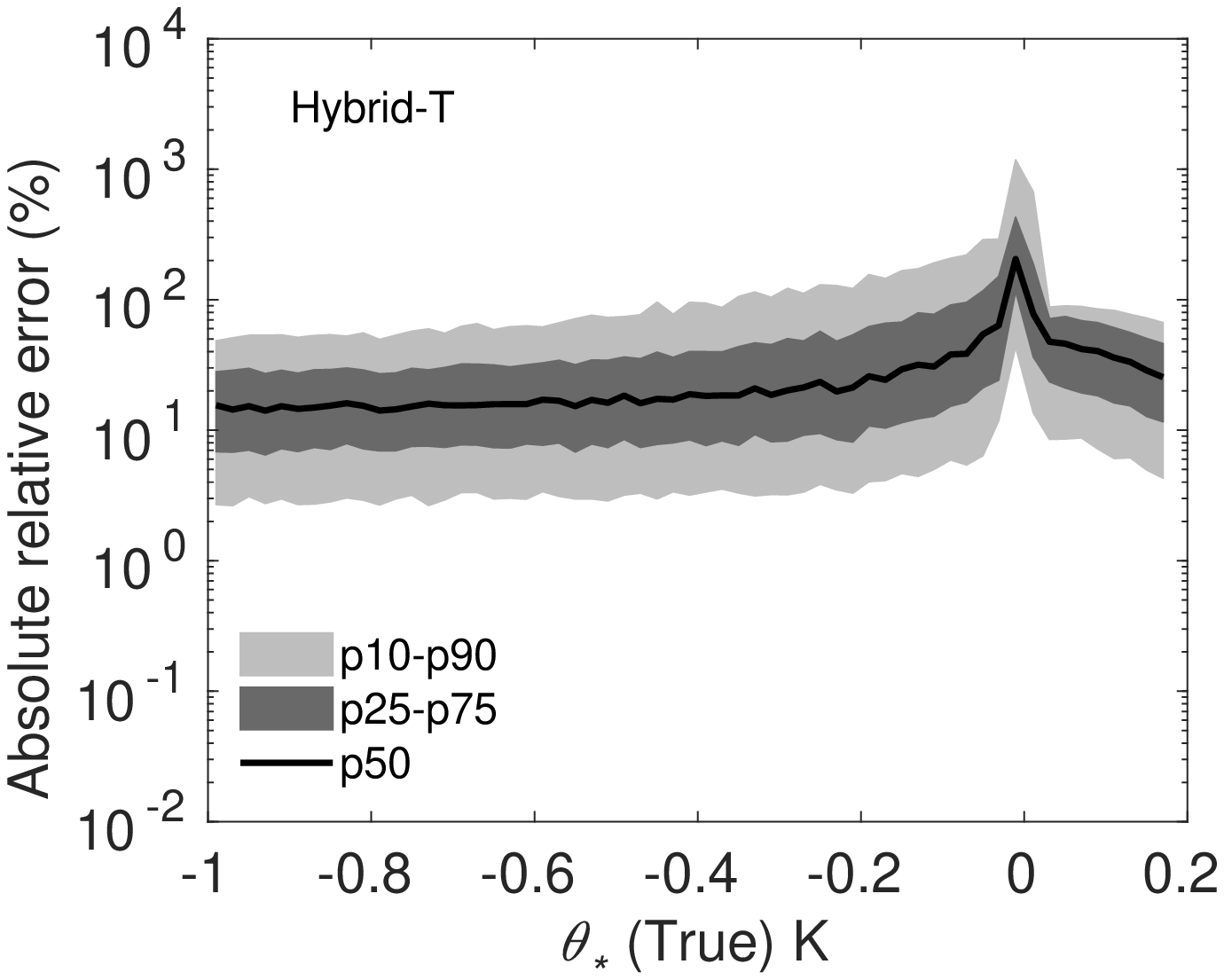}
\caption{Absolute relative errors in the estimation of $u_*$ (top panels) and $\theta_*$ (bottom panels) for two scenarios: 5 (left panels), and 6 (right panels). Flux-estimation approach: hybrid-T.}
\label{HyT}       
\end{figure*}

\begin{figure*}[ht]
\centering
\includegraphics[width=0.49\textwidth]{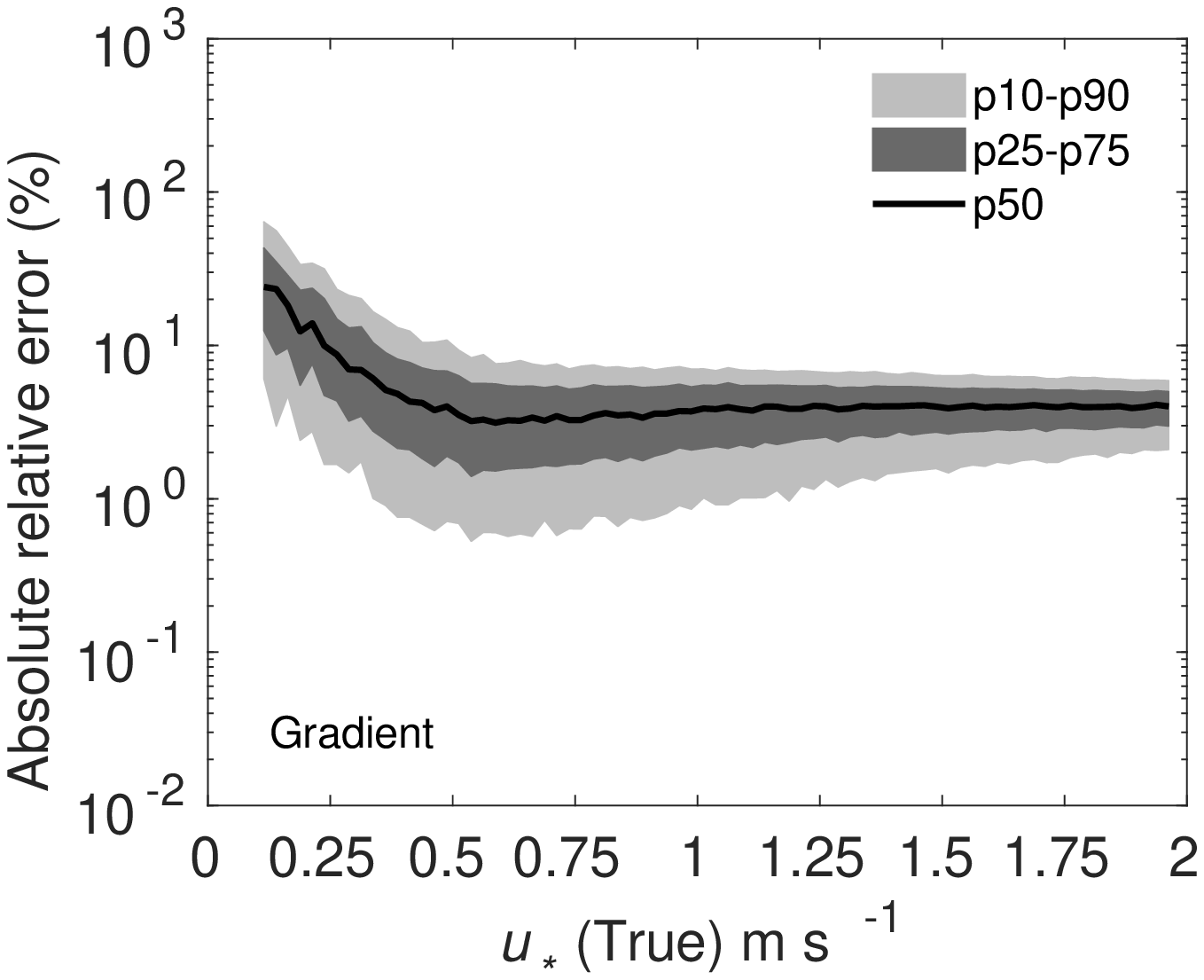}
\includegraphics[width=0.49\textwidth]{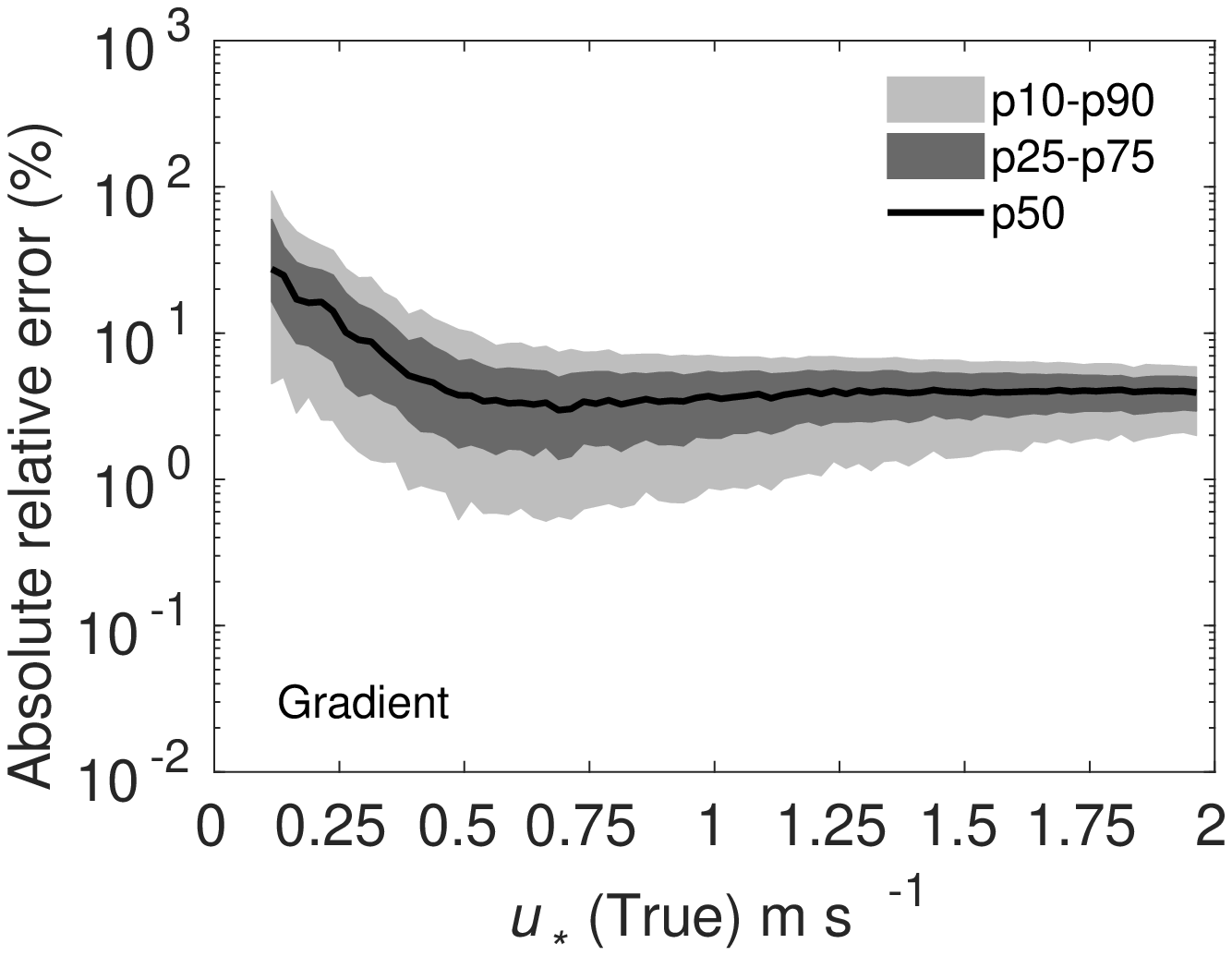}\\
\vspace{0.1in}
\includegraphics[width=0.49\textwidth]{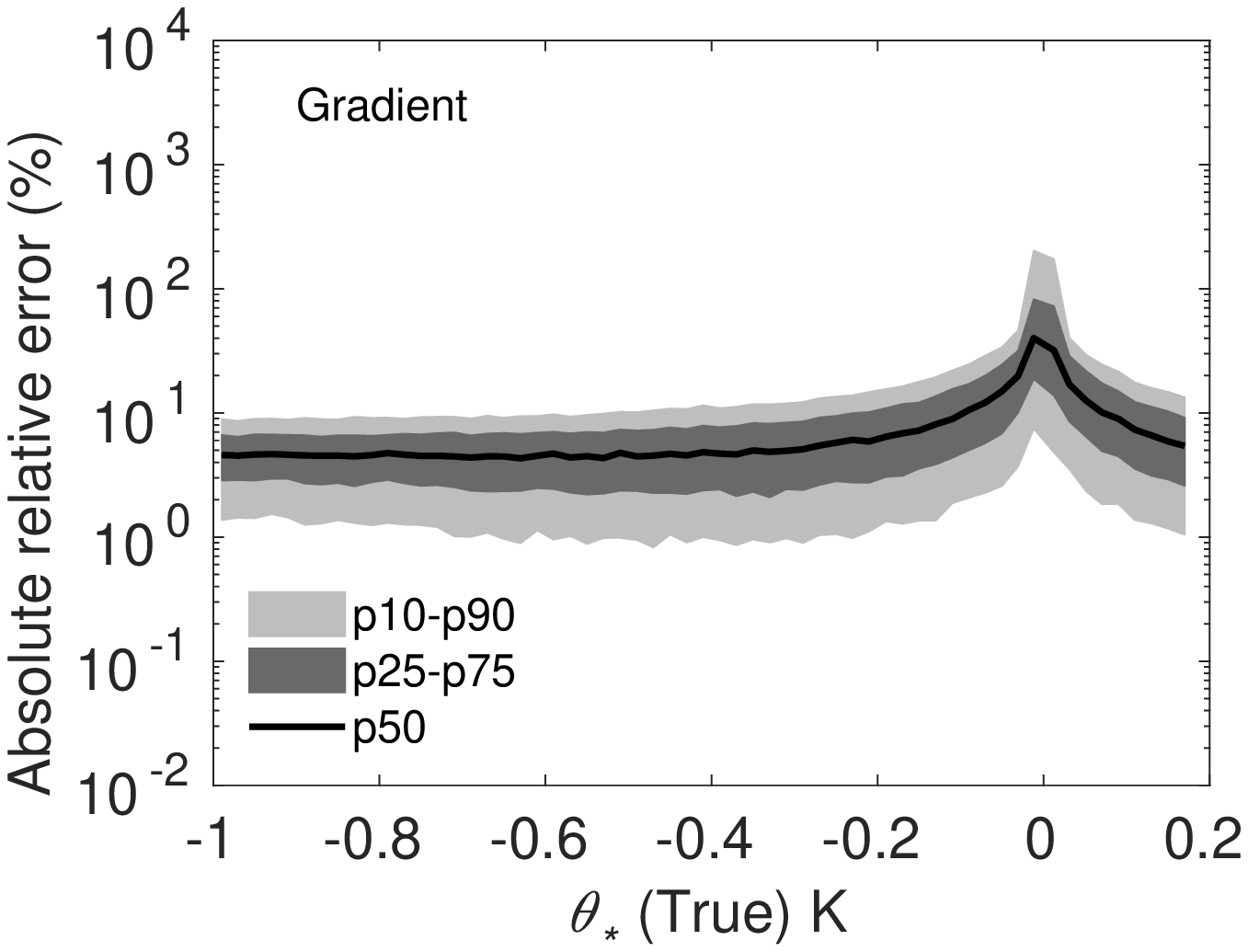}
\includegraphics[width=0.49\textwidth]{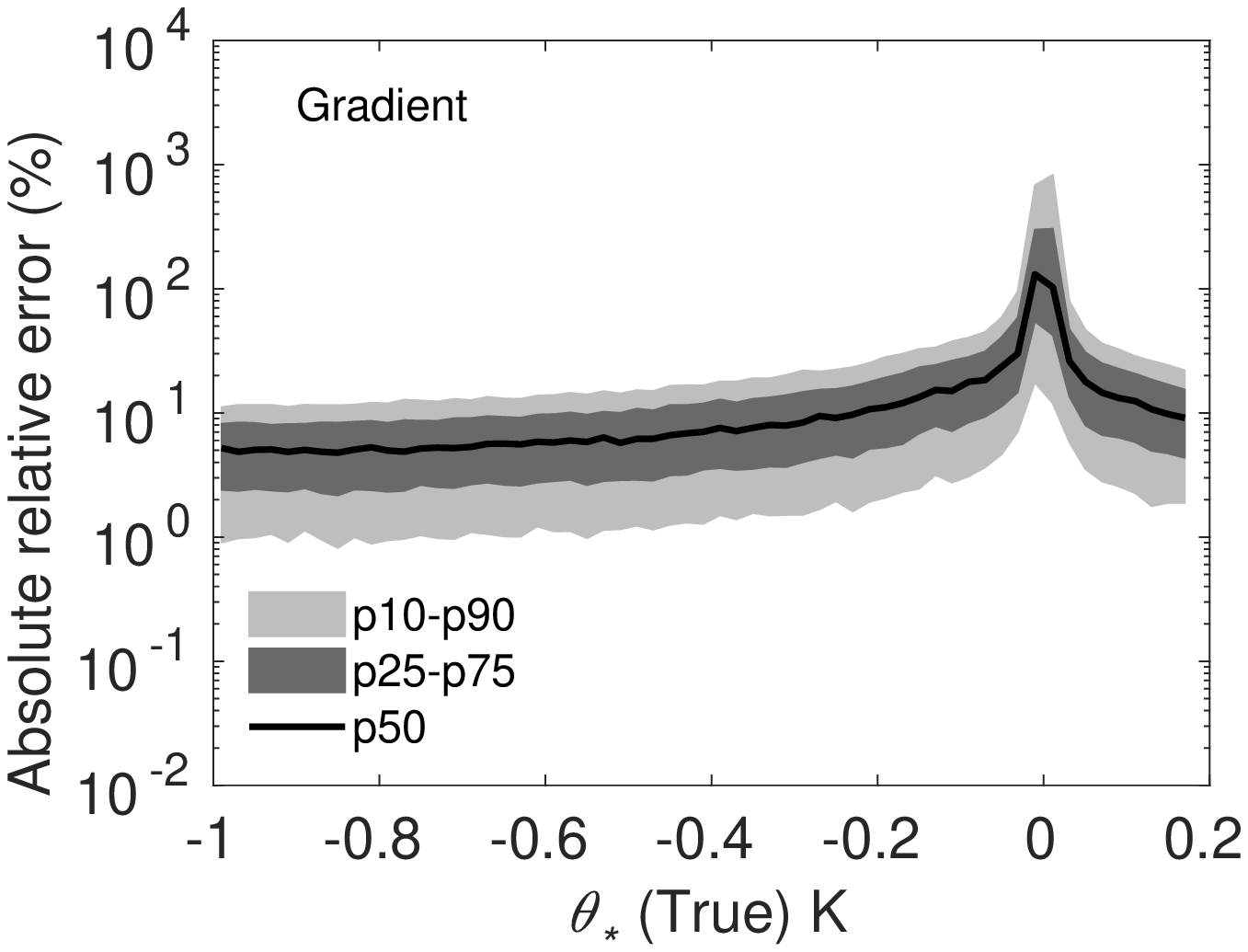}
\caption{Absolute relative errors in the estimation of $u_*$ (top panels) and $\theta_*$ (bottom panels) for two scenarios: 5 (left panels), and 6 (right panels). Flux-estimation approach: traditional gradient method.}
\label{Grad}       
\end{figure*}

\begin{figure*}[ht]
\centering
\includegraphics[width=0.49\textwidth]{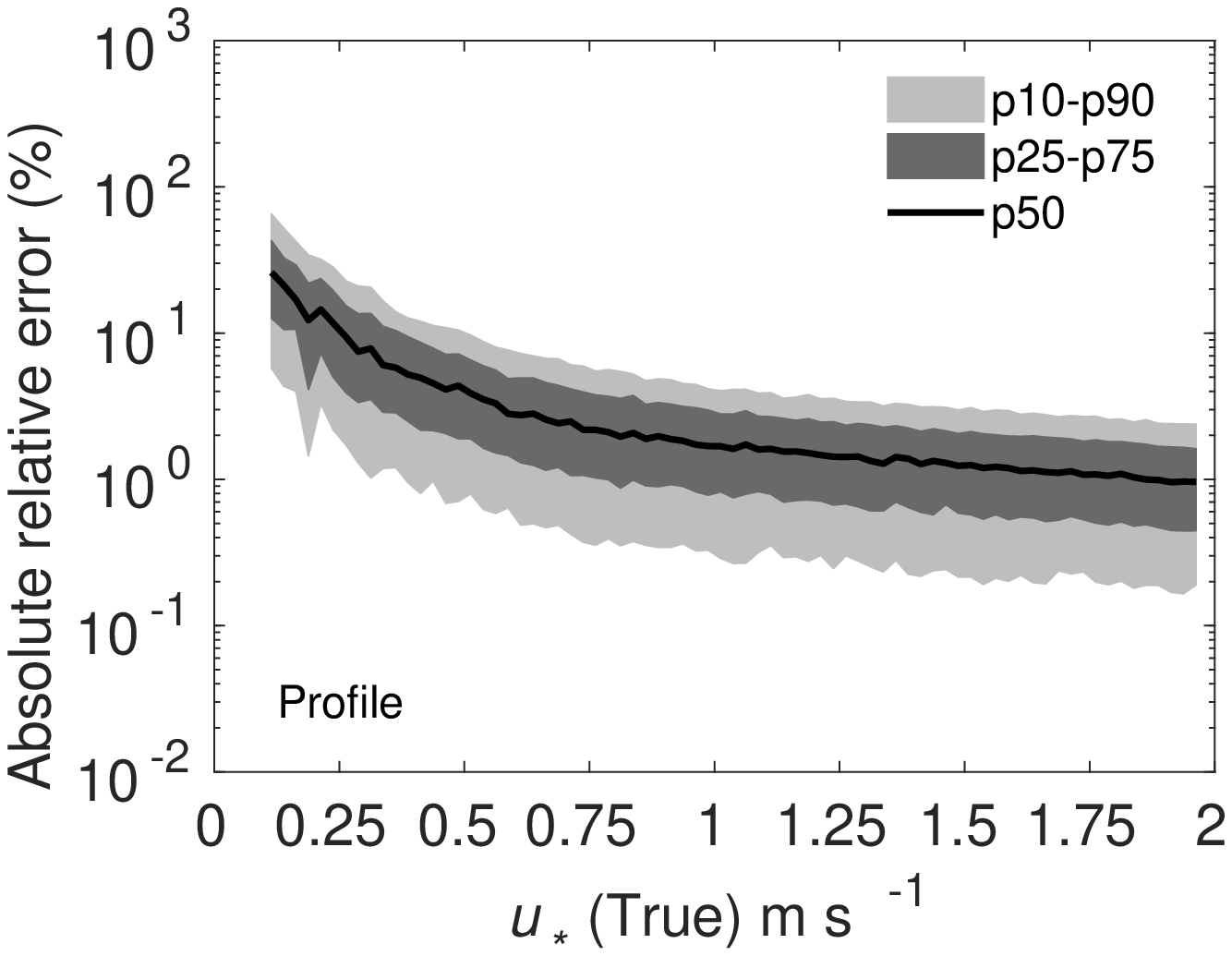}
\includegraphics[width=0.49\textwidth]{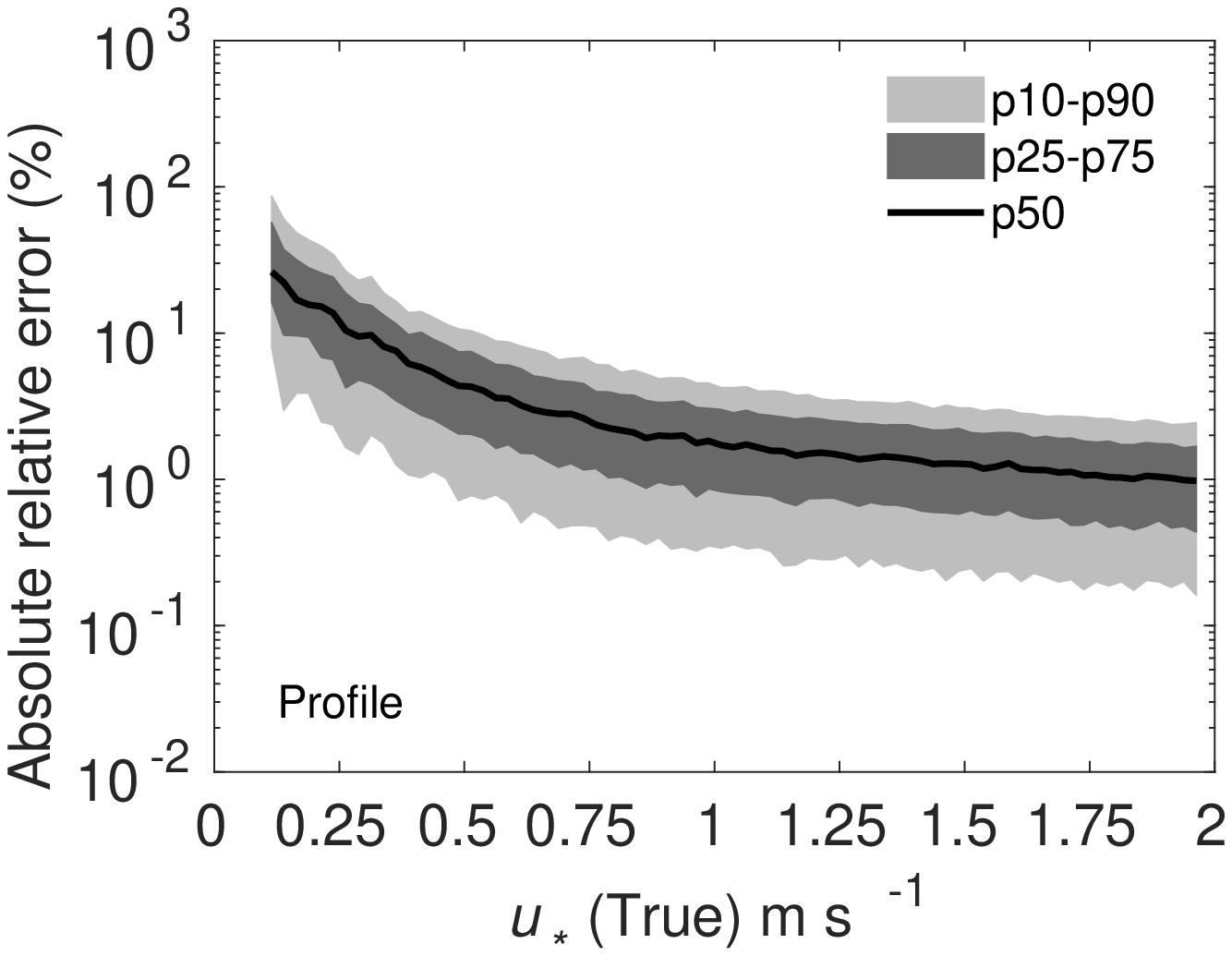}\\
\vspace{0.1in}
\includegraphics[width=0.49\textwidth]{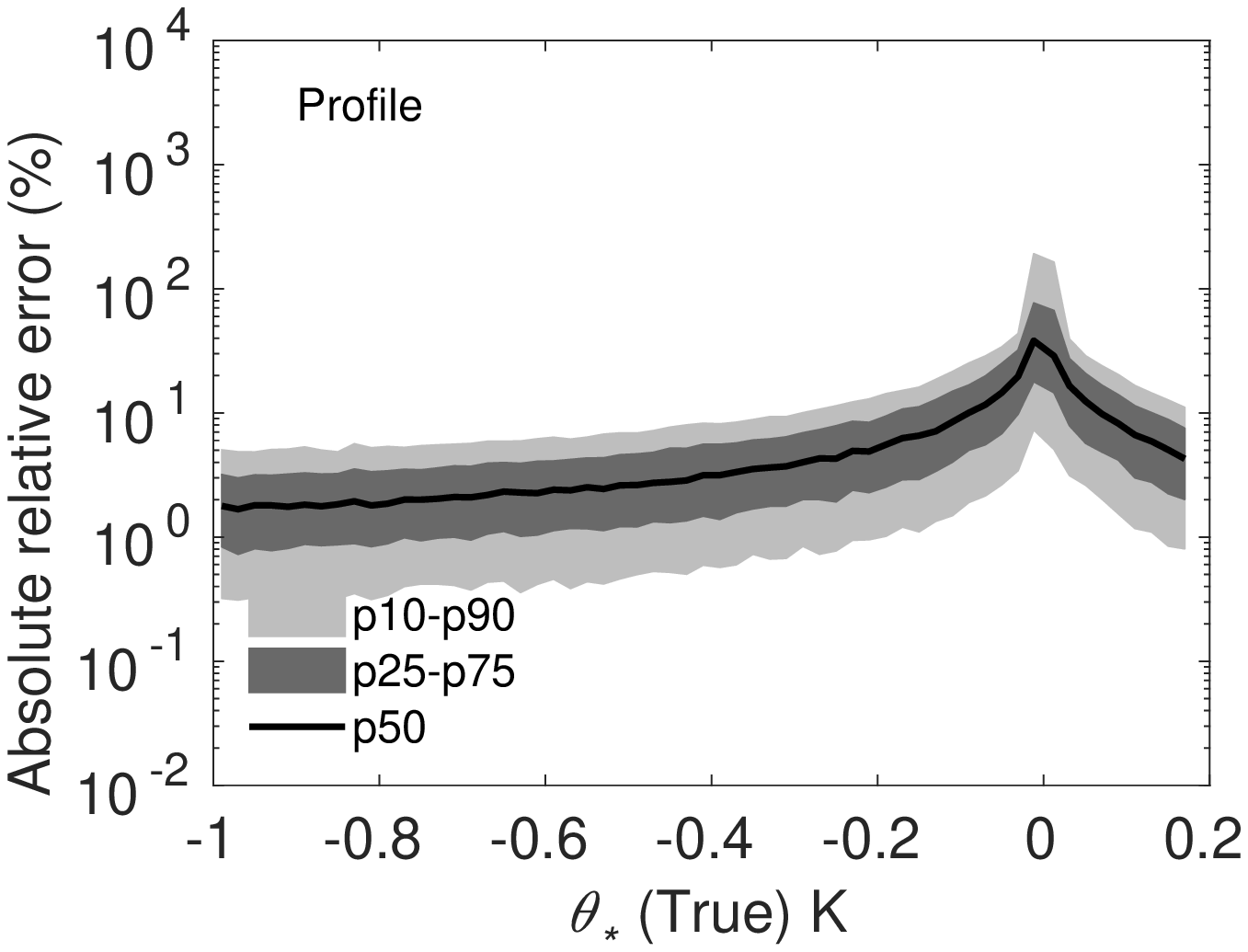}
\includegraphics[width=0.49\textwidth]{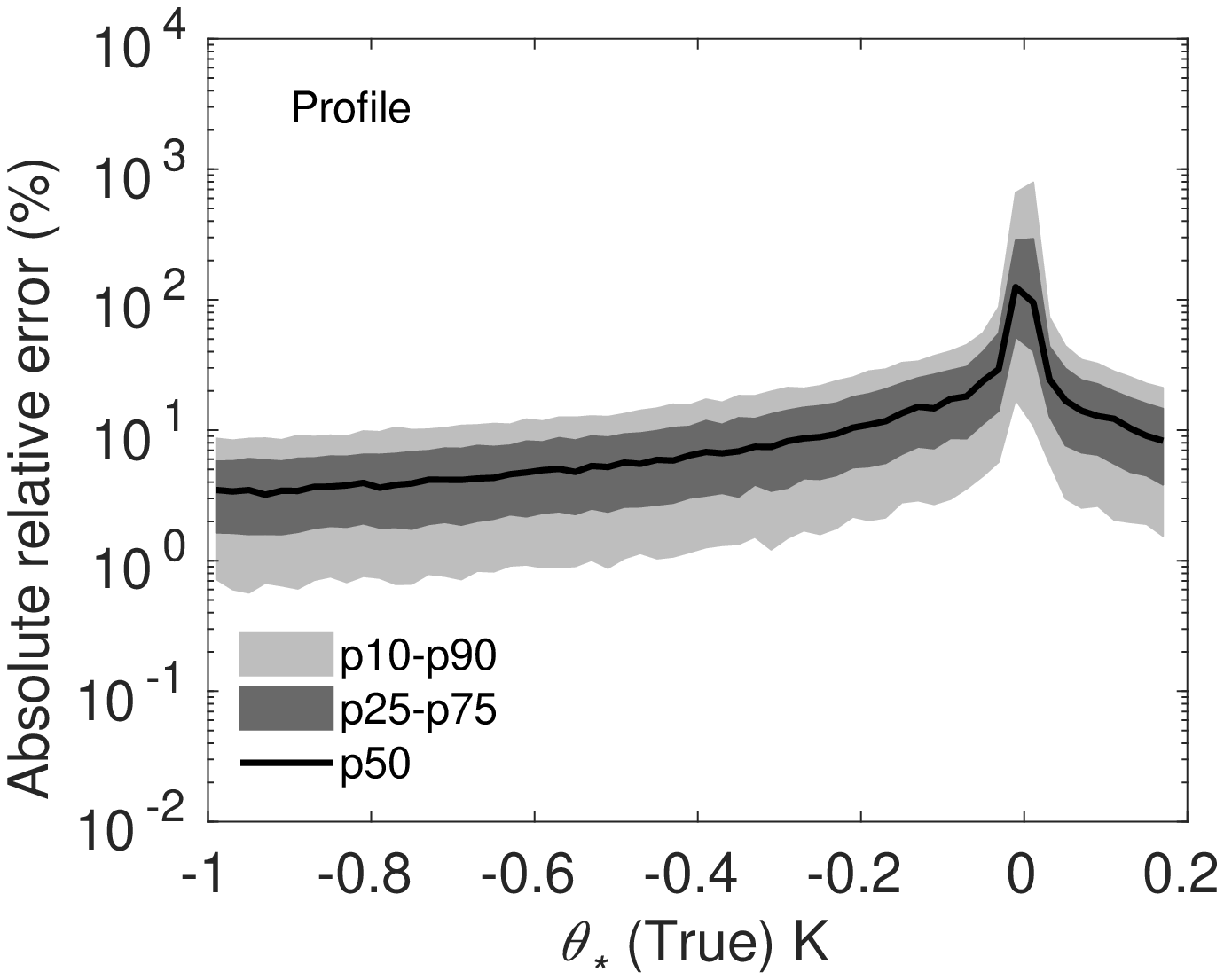}
\caption{Absolute relative errors in the estimation of $u_*$ (top panels) and $\theta_*$ (bottom panels) for two scenarios: 5 (left panels), and 6 (right panels). Flux-estimation approach: traditional profile method.}
\label{Prof}       
\end{figure*}

The results from our uncertainty propagation experiments are shown in Figs.~\ref{HyW_ustar} to \ref{Prof}. In these figures, we report various percentiles of absolute relative errors for both $u_*$ and $\theta_*$. The summary of our results is as follows: 
\begin{itemize}
\item Hybrid-W: for scenarios 1 and 2, the errors in $u_*$ estimation is less than 10\%. However, the errors increase substantially for scenarios 3 and 4. For low $u_*$ values, the errors can range from 10-100\%; however, for high $u_*$ values, they are mostly less than 10\%. The performance of this approach for $\theta_*$ estimation is somewhat poorer. For stable conditions, the median absolute error values are largely on the order of 10-20\%. For unstable conditions, they are higher and seem to be independent of $\theta_*$ values. For near-neutral conditions, large errors can occur due to the division by small numbers. 
\item Hybrid-T: the estimation of $\theta_*$ is far better than $u_*$ for both scenarios 5 and 6. For unstable conditions, the median error values in $\theta_*$ are largely less than 20\%. Marginally higher errors are noticeable for stable conditions. 
\item Gradient: for scenarios 5 and 6, for low $u_*$ values, the errors could be on the order of 10-100\%. Otherwise, for high $u_*$ values, they are much lower than 10\%. For all conditions (with the exception of near-neutral), $\theta_*$ errors are less than 10\%. 
\item Profile: similar to the noise-free cases, this approach outperforms others in both the scenarios 5 and 6. Qualitatively, the errors in $u_*$ estimation follow similar trend as the hybrid-W approach. However, the magnitude of the errors are much smaller. The errors in the estimation of $\theta_*$ also barely exceed 10-20\% (other than the near-neutral conditions). 
\end{itemize}

Before closing, we want to stress that our findings from these uncertainty propagation experiments should be used with caution. We selected specific types of additive noise which are correlated across different heights. Other alternatives are also possible. For example, we used fixed $\sigma$ value for a given scenario; instead, one could use $\sigma$ dependent on the magnitude of $U$ or $\Theta$. In that case, the trends reported in Figs.~\ref{HyW_ustar} to \ref{Prof} would be significantly different. Furthermore, we assumed that the noise in wind-speed and potential temperature profiles are uncorrelated; we do not know if this assumption is realistic or not. In general, high wind speeds lead to lower temperature measurement (radiation) errors; thus, the random errors in wind speeds and temperature might be (anti) correlated.     

\section{Concluding Remarks}
\label{Conclusion}

We have developed new approaches to estimate surface fluxes utilizing either wind-speed or temperature profile data. We have compared our approaches against traditional gradient and profile methods that require both wind-speed and temperature profile data. For noise-free input data, the hybrid approaches perform as well as the traditional profile method. However, in the presence of random errors in input data, the proposed approaches lead to somewhat more flux-estimation errors than the traditional ones. 

Given the unique one-to-one relationships between the ratio of wind-speed differences (or the ratio of potential temperature differences) with the Obukhov length, we propose that either of these ratios could be utilized as a proxy for atmospheric stability. In Basu~(2018)\cite{Basu18}, we demonstrated that the ratio of wind-speed differences was able to categorize observational data in a physically meaningful way. However, further direct verifications are needed.  

We believe that the hybrid-W approach is ideally suited for sodar and lidar-based wind-speed measurements owing to their high vertical resolution in the surface layer. Similarly, the distributed temperature sensing-based high-resolution temperature profiles can be utilized as inputs for the hybrid-T approach. In our future work, observational datasets from various field campaigns will be utilized to make an in-depth assessment of the proposed hybrid-W and hybrid-T approaches. Of course, we will pay close attention to the issues of non-stationarity and heterogeneity, as under such circumstances, the usage of the proposed hybrid approaches (and MOST in general) is not appropriate.   

\section*{Appendix 1: Traditional Gradient and Profile Methods}
\label{Appendix1}

In the traditional gradient method, the following normalized gradient equations are solved in a coupled and iterative manner \citep{Arya01}, 
\begin{subequations}
\begin{equation}
\left(\frac{\kappa z}{u_*}\right)\left(\frac{\partial U}{\partial z}\right) = \phi_m\left(\frac{z}{L} \right),
\end{equation}
\begin{equation}
\left(\frac{\kappa z}{\theta_*}\right)\left(\frac{\partial \Theta}{\partial z}\right) = \phi_h\left(\frac{z}{L} \right).
\end{equation}
\end{subequations}
The vertical gradients are approximated by the finite-difference formulation as follows: \\
$\frac{\partial U}{\partial z} \approx \frac{\Delta U}{\Delta z} = \frac{U(z_2) - U(z_1)}{\left(z_2 - z_1\right)}$, and $\frac{\partial \Theta}{\partial z} \approx \frac{\Delta \Theta}{\Delta z} = \frac{\Theta(z_2) - \Theta(z_1)}{\left(z_2 - z_1\right)}$.\\ 
The estimated gradients are applicable at the mid-point height $z_m = \frac{z_1+z_2}{2}$. Even though this approach (based on linear approximation) is the most popular, an alternative approach utilizing logarithmic approximation was proposed by Arya\cite{Arya91}. For unstable (stable) conditions, the logarithmic (linear) approximation-based approach was found to outperform its counterpart. 

Application of the profile method typically requires the following variables as input: wind speed at one level, temperature at two levels, and aerodynamic roughness length \citep{Berkowicz82}. In a slightly modified version, one uses wind-speed from an additional level instead of the roughness length. One then utilizes the MOST-based profile equations and solves for the unknown fluxes. Brotzge et al.\cite{Brotzge00} utilized this modified profile approach to estimate fluxes from the Oklahoma mesonet.   

\section*{Appendix~2: Swinbank's Exponential Wind Profile}
\label{Appendix2}

Swinbank\cite{Swinbank64} proposed the following equation for surface-layer wind profile,  
\begin{equation}
\label{Swinbank1}
U\left(z_2\right) - U\left(z_1\right) = \frac{u_*}{\kappa}\ln\left[\frac{\exp\left(\frac{z_2}{L}\right) - 1}{\exp\left(\frac{z_1}{L}\right) - 1}\right],
\end{equation}
and further derived, 
\begin{equation}
\label{Swinbank2}
\frac{U\left(z_3\right) - U\left(z_1\right)}{U\left(z_2\right) - U\left(z_1\right)} = \frac{\ln\left[\frac{\exp\left(\frac{z_3}{L}\right) - 1}{\exp\left(\frac{z_1}{L}\right) - 1}\right]}{\ln\left[\frac{\exp\left(\frac{z_2}{L}\right) - 1}{\exp\left(\frac{z_1}{L}\right) - 1}\right]}.
\end{equation}
commenting that Eq.~\ref{Swinbank2} permits the determination of $L$ from observed wind-speed data at three levels using numerical or graphical interpolation. Once $L$ is determined, $u_*$ can be estimated from Eq.~\ref{Swinbank1}. Our proposed hybrid-W approach is almost identical, albeit  it makes use of Eq.~\ref{RatioW}. 

\section*{Appendix 3: Stability Correction Functions}
\label{Appendix3}

Over the years, numerous stability correction functions have been proposed in the literature. A few of them are listed below: \\

\noindent \textbf{Dyer and Hicks\cite{Dyer70}, Businger et al.\cite{Businger71}, Dyer\cite{Dyer74}:}

\begin{subequations}
\begin{equation}
\psi_m = 2\ln\left(\frac{1+x}{2} \right) + \ln\left( \frac{1+x^2}{2} \right) - 2\tan^{-1}x + \frac{\pi}{2}; \hspace{0.2in} \mbox{for } \frac{z}{L} \le 0
\label{psimBusingerU}
\end{equation}
\begin{equation}
\psi_h = 2\ln\left(\frac{1+x^2}{2} \right); \hspace{0.2in} \mbox{for } \frac{z}{L} \le 0
\label{psihBusingerU}
\end{equation}
\begin{equation}
\psi_m = \psi_h = -5\frac{z}{L}; \hspace{0.2in} \mbox{for } \frac{z}{L} \ge 0
\label{psiBusingerS}
\end{equation}
\label{psiBusinger}
\end{subequations}
where $x = \left( 1 - 16\frac{z}{L}\right)^{1/4}$.\\

\noindent \textbf{Beljaars and Holtslag\cite{Beljaars91}:}

\begin{subequations}
\begin{equation}
\psi_m = -a\frac{z}{L} -b\left(\frac{z}{L} - \frac{c}{d} \right)\exp\left(-d\frac{z}{L} \right) -\frac{bc}{d}; \hspace{0.2in} \mbox{for } \frac{z}{L} \ge 0
\label{psimBeljaarsS1}
\end{equation}
\begin{equation}
\psi_h = -\left(1 + \frac{2a}{3}\frac{z}{L} \right)^{3/2} -b\left(\frac{z}{L}-\frac{c}{d}\right)\exp \left(-d\frac{z}{L} \right) -\frac{bc}{d}+1; \hspace{0.2in} \mbox{for } \frac{z}{L} \ge 0
\label{psimBeljaarsS2}
\end{equation}
\end{subequations}
where $a = 1$, $b = \frac{2}{3}$, $c = 5$, and $d = 0.35$.\\

\noindent \textbf{Duynkerke\cite{Duynkerke91}:}

\begin{subequations}
\begin{equation}
\psi_m = -\left(1+ \frac{\beta_m}{\alpha_m}\frac{z}{L}\right)^{\alpha_m}; \hspace{0.2in} \mbox{for } \frac{z}{L} \ge 0
\label{psimDuynkerkeS}
\end{equation}
\begin{equation}
\psi_h = -\left(1+ \frac{\beta_h}{\alpha_h}\frac{z}{L}\right)^{\alpha_h}; \hspace{0.2in} \mbox{for } \frac{z}{L} \ge 0
\label{psihDuynkerkeS}
\end{equation}
\end{subequations}
where $\alpha_m = \alpha_h = 0.8$, $\beta_m = 5$, and $\beta_h = 7.5$. \\

\noindent \textbf{Wilson\cite{Wilson01}:}

\begin{subequations}
\begin{equation}
\psi_m = 3\ln \left( \frac{1+\sqrt{1+\gamma_m |z/L|^{2/3}}}{1+\sqrt{1+\gamma_m |z_0/L|^{2/3}}} \right); \hspace{0.2in} \mbox{for } \frac{z}{L} \le 0
\end{equation}
\begin{equation}
\psi_h = 3\ln \left( \frac{1+\sqrt{1+\gamma_h |z/L|^{2/3}}}{1+\sqrt{1+\gamma_h |z_{0 T}/L|^{2/3}}} \right); \hspace{0.2in} \mbox{for } \frac{z}{L} \le 0
\end{equation}
\end{subequations}
where $\gamma_m = 3.6$ and $\gamma_h = 7.9$. \\

\noindent \textbf{Cheng and Brutsaert\cite{Cheng05}:}

\begin{subequations}
\begin{equation}
\psi_m = -a\ln \left(\frac{z}{L} + \left(1 + \left(\frac{z}{L}\right)^b\right)^{1/b}\right); \hspace{0.2in} \mbox{for } \frac{z}{L} \ge 0
\end{equation}
\begin{equation}
\psi_h = -c\ln \left(\frac{z}{L} + \left(1 + \left(\frac{z}{L}\right)^d\right)^{1/d}\right); \hspace{0.2in} \mbox{for } \frac{z}{L} \ge 0
\end{equation}
\end{subequations}
where $a, b, c,$ and $d$ equal to 6.1, 2.5, 5.3, and 1.1, respectively.

\begin{acknowledgements}
The author is grateful to Fred Bosveld, Stephan de Roode, Bert Holtslag, Harm Jonker, Branko Kosovi\'{c}, Peggy LeMone, Larry Mahrt, Pier Siebesma, and Bas van de Wiel for their constructive feedback on this work. 
\end{acknowledgements}

\bibliographystyle{spbasic}      

\end{document}